\documentclass[11pt]{JHEP3}
\usepackage{amssymb}
\usepackage{amsmath}
\usepackage{amsfonts}
\usepackage{epsfig}

\author{Liat Maoz, Vyacheslav S. Rychkov \\
\ \\
Institute for Theoretical Physics, University of Amsterdam \\
Valckenierstraat 65, 1018XE Amsterdam, The Netherlands\\
\ E-mail: \email{lmaoz@science.uva.nl},
\email{rychkov@science.uva.nl}}
\title{Geometry Quantization from Supergravity:\\ the case of ``Bubbling AdS''}

\preprint{ITFA-2005-38} \abstract{We consider the moduli space of
1/2 BPS configurations of type IIB SUGRA found by Lin, Lunin and
Maldacena (hep-th/0409174), and quantize it directly from the
supergravity action, around any point in the moduli space. This
quantization is done using the Crnkovi\'{c}-Witten-Zuckerman
covariant method. We make some remarks on the applicability and
validity of this general on-shell quantization method. We then
obtain an expression for the symplectic form on the moduli space
of LLM configurations, and show that it \emph{exactly} coincides
with the one expected from the dual fermion picture. This
equivalence is shown for any shape and topology of the droplets
and for any number of droplets. This work therefore generalizes
the previous work (hep-th/0505079) and resolves the puzzle
encountered there.}
\begin{document}
\section{Introduction}

One of the greatest achievements of string theory is the fact that
it is able to explain gravity, or supergravity, in terms of a
microscopic set of degrees of freedom --- oscillation modes of
strings with various boundary conditions.

Undoubtedly, if string theory is to be a quantum theory of gravity, then
such a description is essential. One particular application of having a
microscopic set of degrees of freedom amenable to quantization is the
ability to count the number of microstates corresponding to a specific
geometry, thus determining its microscopic entropy. One of the most
fascinating applications of this is for geometries describing black holes.

In general a coherent state of string oscillations generates a
supergravity background. However, there are very few such
backgrounds for which we know the corresponding coherent state of
string oscillations. In most cases, given some supergravity
background, it will be very hard to find the exact coherent state
making it up. In some specific cases, when the spacetime is
asymptotically AdS, one could use AdS/CFT \cite{AdSCFT} to obtain
a different description of the supergravity configuration --- in
terms of microscopic degrees of freedom living in a conformal
field theory, or in a deformation thereof. Such a description
requires a good knowledge of the boundary theory and of the
bulk/boundary dictionary, which is not necessarily available.

It is therefore clear that it would be very useful if one could
quantize geometries directly from the supergravity picture,
without recourse to a different microscopic description of the
system. In this paper, continuing the previous work done in
collaboration with L.~Grant, J.~Marsano and K.~Papadodimas in
\cite{our}, we develop a method to do precisely that, which we
propose to call \emph{on-shell quantization}.

As the name suggests, on-shell quantization is not a method to quantize all
possible fluctuation modes of supergravity, but is used only to quantize a
subspace of fluctuations all describing on-shell SUGRA configurations within
a given moduli space. In this sense it is a special case of mini-superspace
quantization \cite{Misner}. The method is only applicable when on the moduli
space of solutions all fluxes are kept fixed.

When applicable, this method is used to quantize only the moduli
space of solutions, freezing all other fluctuations of the fields.
Thus one might wonder how meaningful the results of such a
quantization are in the general framework. In general indeed one
should expect corrections to the results of the on-shell
quantization, and one can try to estimate their size. It is only
in particular cases where the moduli space of solutions is also
protected by some symmetry, that this sector could completely
decouple from the rest of the Hilbert space, and the energy
spectrum computed by the on-shell quantization method would remain
uncorrected.

In this paper we apply the on-shell quantization method to a certain moduli
space of Type IIB SUGRA, which both has fixed fluxes and is also protected
by supersymmetry, so that the results of the on-shell quantization remain
uncorrected. This is the set of 1/2 BPS solutions found by Lin, Lunin and
Maldacena in \cite{LLM}. The LLM solutions are in one-to-one correspondence
with various 2-colorings of a 2-plane, usually referred to as `droplets'.
The deformations of the droplet shapes such that the areas of different
white and black regions remain fixed are exactly the deformations which
leave the fluxes fixed. Thus what we would like to do is to quantize the
moduli space of such fluctuations around a general droplet configuration%
\footnote{%
Note that we strive to have a quantization which relies only on SUGRA and
not on some microscopic string or brane picture, such as was done for
instance in \cite{Mandal} or, in a different context, in \cite{Marolf}. It
should be noted that although some remarks on the problem of performing a
SUGRA quantization were made in \cite{Mandal}, they seem to apply only in a
partial case when there is a preferred group action on the moduli space of
solutions. In this paper we follow a different, more direct, path.}.

The LLM solutions have the feature that if the droplets are of finite size,
then the spacetime is asymptotically $AdS_{5}\times S^{5}$. As we remarked
before, this asymptotic behavior enables one to use AdS/CFT and get an
equivalent description in terms of the $\mathcal{N}=4$ SYM field theory. The
specific sector of 1/2 BPS solutions considered by LLM turns out to be a
subsector of SYM, admitting description in terms of free fermions in a
harmonic oscillator potential \cite{Corley,Berenstein}. In fact the droplet
describing the supergravity solution turned out to be the same droplet in
the phase space describing the free fermions. Thus this is one of the
special cases where string theory provides us with a dual description of the
system where microscopic degrees of freedom are manifest. It is the Hilbert
space of these free fermions in a harmonic oscillator, which we would like
to derive entirely from the SUGRA picture using the on-shell quantization.

In a previous paper with L.~Grant, J.~Marsano and K.~Papadodimas \cite{our},
we gave general expressions for the symplectic form on the moduli space of
LLM solutions, and applied them to two specific backgrounds: $AdS_{5}\times
S^{5}$ and the plane wave. In the case of $AdS_{5}\times S^{5}$ we
completely reproduced the fermion symplectic form from the SUGRA symplectic
currents. In the plane wave case we found the symplectic form of the same
functional expression as we expect from the fermion analysis, however the
numerical coefficient was a factor of 2 smaller than expected. In that paper
we gave a few speculations on the source of that discrepancy.

In this paper we resolve the puzzle. We make the observation that one must
work in a gauge where the variations of the SUGRA gauge fields are
everywhere regular. Different gauge choices result in extra boundary terms
in the symplectic current, and thus working in a singular gauge, as we did
in \cite{our} might result in missing terms in the symplectic form. In this
paper we succeed in evaluating the SUGRA symplectic form around \emph{any of
the LLM backgrounds}, finding the expected fermion symplectic form with
exact numerical matching.

This paper is organized as follows. In section 2 we first discuss the
general issue of quantizing geometries from the supergravity action. We
explain the motivation and main ideas of the 'on-shell quantization' method,
and show how to apply it to pure gravity as well as to general Lagrangian
theories. A central role in this discussion is played by the \textit{CWZ
symplectic currents}, introduced 20 years ago by Crnkovi\'{c} and Witten
\cite{CW} and by Zuckerman \cite{Zuckerman}. We then turn to discuss when
this method is applicable, and how one may expect it to be corrected by
quantization of the full theory.

Then in section 3 we apply the general idea and expressions to the case of
the "Bubbling AdS" geometries found by LLM \cite{LLM}. We first analyze the
dual fermion system and find an expression for the symplectic form in the
large $N$ limit. This is the expression we aim to independently derive from
the supergravity analysis. Then we indeed turn to the moduli space of
supergravity solutions, and show that using the CWZ currents we are able to
reproduce this form, first in the specific case of the plane wave geometry,
and then in the most general droplet case. We end in section 4 with
conclusions and directions for future research. Some technical details are
deferred to the appendix.

Results of this paper were reported in a talk presented by the second author
at Strings 2005, Toronto.

\section{Geometry quantization from supergravity}

\label{2}

\subsection{Motivation}

\label{2.1}

Given a continuous family of supergravity solutions, it is natural to ask
how the moduli space of this family is going to be quantized. A
well-understood special case of such quantization is the Dirac quantization
condition imposed on the fluxes of gauge fields present in the theory:
typically, such fluxes $\mathcal{F}$ on closed cycles have to be quantized
in units of an elementary flux: $\mathcal{F}=\mathcal{F}_{0}n$, $n\in\mathbb{%
Z}$.

However, sometimes the moduli space will contain deformations corresponding
to all fluxes kept fixed. In this paper we are mostly concerned with how to
quantize the moduli space in such a situation, which is in a sense
complementary to the flux quantization.

As a characteristic example, consider the family of D1-D5 solutions with
angular momentum found in \cite{LMM}. The moduli space of solutions is
parametrized by 4-dimensional closed curves; there are no nontrivial fluxes.
This family plays an important role in Mathur's program of describing black
hole microstates by regular geometries, providing microstates for the
2-charge D1-D5 black hole \cite{Mathur}\footnote{%
In \cite{Marika} another family of SUGRA solutions is constructed, which is
supposed to represent additional microstate geometries.}.

Another interesting recent example is the LLM family of Type IIB\ solutions
with $AdS_{5}\times S^{5}$ asymptotics \cite{LLM}\footnote{%
LLM \cite{LLM} also found similar M-theory solutions with $AdS_{4}\times
S^{7}$ and $AdS_{7}\times S^{4}$ asymptotics, and other families of
solutions have been generated subsequently by applying the LLM method to
different theories in various dimensions \cite{Diana},\cite{Martelli},\cite%
{Chong}. These families became collectively known as \textquotedblleft
Bubbling AdS".}. The moduli space is parametrized by planar droplets of
various shapes. There is an infinite-dimensional family of deformations
within the moduli space corresponding to keeping the fluxes fixed: these are
the deformations keeping fixed areas of all black and white regions. We will
discuss these solutions and their moduli space quantization in detail in
Section \ref{3}.

For both of the above families, there is a dual description --- free
fermions in the LLM case, chiral fundamental string excitations in the D1-D5
case --- which can be used to quantize the moduli space. These dual
descriptions can be derived by using AdS/CFT or various other indirect
methods (giant graviton picture for LLM, string dualities for D1-D5).
However, as we will explain below, there exists a method to derive such
quantization results directly from supergravity. In some cases this might be
the only way to quantize SUGRA systems, since a dual microscopic description
is not always known.

Counting 3-charge D1-D5-P black hole microstate geometries in the context of
Mathur's program could provide an example of a situation when our direct
method may lead to progress which will otherwise be difficult to achieve. So
far only some specific families of regular 3-charge geometries were
constructed \cite{Lunin}. It is conjectured that the general case can be
understood in terms of supertubes \cite{bena0}, however a world-volume
description of the 3-charge supertube configurations, which could be used to
quantize them, is not yet known\footnote{%
There has also been some other work trying to identify CFT states with
particular 3-charge D1-D5-P geometries \cite{benacft}.}. In that case our
direct method may be the only way to quantize and count these geometries,
once they are fully described.

\subsection{General idea}

\label{2.2} The general idea of the method is simple. Every supergravity
theory, as any Lagrangian theory, comes equipped with a symplectic form $%
\Omega$, which is defined on the full phase space of this theory. The given
moduli space of solutions $\mathcal{M}$, which we would like to quantize,
forms a subspace of the full phase space. All we have to do is to restrict $%
\Omega$ to $\mathcal{M}$, which will give us the symplectic form on $%
\mathcal{M}$:
\begin{equation}
\omega=\Omega|_{\mathcal{M}}\,.  \label{restrict}
\end{equation}
Once $\omega$ is found, it can be quantized in the usual way. We will call
this method \textit{on-shell quantization}\footnote{\textit{\ }We called it
`minisuperspace quantization' in \cite{our}, however this term is used in a
slightly wider sense in the canonical gravity literature, implying a
reduction of degrees of freedom (typically by imposing a symmetry), but not
necessarily explicit knowledge of all solutions.}.

This philosophy is completely general and applies to any Lagrangian theory.
For example, let us quantize the chiral sector of a 2-dimensional free boson
using this method. The action is
\begin{equation}
S=\frac{1}{2}\int dt\,dx\left( \dot{\phi}^{2}-\phi^{\prime2}\right) \,,
\end{equation}
and the symplectic form is given by the familiar expression%
\begin{equation}
\Omega=\int_{t=\text{const}}dx\,\delta\pi(t,x)\wedge\delta\phi(t,x),\qquad
\pi=\dot{\phi}.  \label{b-form}
\end{equation}
This is a 2-form on the full phase space of the theory (which can be thought
of as the space of solutions of equations of motion). The 1-forms $%
\delta\phi $ and $\delta\pi$ (where $\delta$ is the exterior derivative
operation) can be thought to represent fluctuations around the solutions. We
can either use the wedge product $\wedge$ as in (\ref{b-form}) or think of
1-forms as anticommuting quantities.

Now we should pick the moduli space of solutions to be quantized, which we
choose to be the space of left-moving solutions, parametrized by an
arbitrary function:%
\begin{equation}
\mathcal{M}=\{\phi(t,x)=f(t+x),\text{ }f\text{ arbitrary}\}.
\end{equation}
It is a simple matter to restrict $\Omega$ to $\mathcal{M}$. We get:%
\begin{equation}
\omega=\Omega|_{\mathcal{M}}\,=\int du\,\delta f^{\prime}(u)\wedge\delta
f(u).
\end{equation}
Now from $\omega$ we can infer the Poisson bracket:%
\begin{equation}
\{f(u_{1}),f^{\prime}(u_{2})\}=\delta(u_{1}-u_{2}).
\end{equation}
This Poisson bracket can then be promoted to a quantum commutator using the
Dirac prescription:
\begin{equation}
\lbrack\hat{f}(u_{1}),\hat{f}^{\prime}(u_{2})]=i\hbar\,\delta(u_{1}-u_{2}).
\end{equation}
The Fock space representation of this commutator is then constructed in the
usual way:%
\begin{align}
\hat{f}(u) & =\int_{0}^{\infty}\frac{dp}{2\pi}\frac{1}{\sqrt{2p}}%
e^{-ipu}\alpha_{p}+h.c.\,, \\
\lbrack\alpha_{p},\alpha_{p^{\prime}}^{\dagger}] & =2\pi\hbar\,\delta
(p-p^{\prime})\,.
\end{align}
This way we recover the standard quantization of the chiral boson sector,
which is equivalent to quantizing the full theory and then putting all
right-movers in their vacuum state. Notice that the only piece of off-shell
information we are using is the full symplectic form (\ref{b-form}). All the
remaining computations are done on shell, that is in $\mathcal{M}$.

\subsection{Symplectic form of gravity}

\label{2.4}

We are interested in theories which contain gravity as a subsector. Thus a
necessary prerequisite is the symplectic form of pure gravity, which we will
now discuss. We will assume that the solutions we have to quantize are
regular, and that an initial value surface, $\Sigma,$ can be chosen. We will
also assume that there are no horizons present\footnote{%
In presence of horizons one would have to consider a Cauchy surface in the
extended black hole spacetime, which may or may not exist, and could contain
additional causally disconnected asymptotic regions. The nature of degrees
of freedom which are being quantized in this case is not clear to us.
Alternatively, one could try to treat horizons as boundaries. Additional
analysis is required to decide which approach is correct. For now we prefer
to exclude this case from discussion.}. All these assumptions will be
satisfied in the examples to be considered below.

The most direct route to the symplectic form of gravity lies via the
canonical formalism. Gravity can be put in the canonical form using the ADM
splitting \cite{ADM} of the metric:%
\begin{equation}
ds^{2}=-N^{2}dt^{2}+h_{ij}(dx^{i}+N^{i}dt)(dx^{j}+N^{j}dt).  \label{ADM}
\end{equation}
The phase space is parametrized by the metric on $\Sigma,$ $h_{ij}$, and by
the corresponding canonically conjugate momenta $\Pi^{ij}.$ These momenta
are related to the extrinsic curvature $K^{ij}$ of the Cauchy surface $%
\Sigma $:%
\begin{align}
\Pi^{ij} & =\sqrt{h}(K^{ij}-K^{l}{}_{l}h^{ij})\,,  \label{momenta} \\
K_{ij} & =\frac{1}{2N}(\dot{h}_{ij}-D_{i}N_{j}-D_{j}N_{i})\,.  \label{curv}
\end{align}
The canonical variables satisfy a set of nonlinear constraints, which have
to be realized as operator relations in quantum theory. This is a highly
nontrivial task when quantizing the full theory. Fortunately, in our case
this won't be a problem: we are quantizing families of classical solutions,
and all constraints will be automatically satisfied.

In this formalism, the symplectic form is given by the natural expression:%
\begin{equation}
\Omega=\int_{\Sigma}d^{D-1}x\,\delta\Pi^{ij}(t,x)\wedge\delta h_{ij}(t,x)\,.
\label{can}
\end{equation}
To restrict this symplectic form to the moduli space of solutions $\mathcal{M%
}$, one first has to put all metrics of the family in the ADM form, and
evaluate the canonical momenta using (\ref{momenta}). This is doable in
principle, but can be rather cumbersome in practice. Fortunately, an
equivalent method exists, which avoids using the ADM split.

\subsection{Covariant approach}

\label{2.5} In the equivalent covariant approach, which is computationally
much simpler than the direct method outlined in the previous subsection, one
expresses the symplectic form as an integral of a \textit{symplectic current}
over the Cauchy surface $\Sigma$:%
\begin{equation}
\Omega=\int d\Sigma_{l}J^{l}\,.  \label{sform}
\end{equation}
The symplectic current of gravity was found by Crnkovi\'{c} and Witten \cite%
{CW} and is given by the following covariant expression containing
variations of the metric and the Christoffel symbols\footnote{%
Notice that \cite{CW} defines the symplectic form as $\int d\Sigma_{l}\sqrt{%
-g}J^{l},$ so that our symplectic currents differ from \cite{CW} by a factor
of $\sqrt{-g}$.}:%
\begin{equation}
J^{l}=-\delta\Gamma_{mn}^{l}\wedge\delta(\sqrt{-g}g^{mn})+\delta\Gamma
_{mn}^{n}\wedge\delta(\sqrt{-g}g^{lm})\,.  \label{CW}
\end{equation}
This current has a number of remarkable properties. First of all,
it is conserved: $\partial_{l}J^{l}=0$, which makes $\Omega$
invariant under variations of $\Sigma$ (local variations in
general, as well as global variations if the metric perturbations
have fast enough decay at infinity). Second, $J^{l}$ changes by a
total derivative if the metric perturbation
undergoes a linear gauge transformation:%
\begin{equation}
\delta g_{mn}\rightarrow\delta g_{mn}+\nabla_{(m}\xi_{n)}\,.
\end{equation}
This property renders $\Omega$ defined by (\ref{sform}) gauge invariant
(assuming that $\xi$ has finite support or decays sufficiently fast at
infinity). It should be noted that both properties are true only if the
metric is varied inside a moduli space of \textit{solutions}, i.e.\ the
background satisfies Einstein's equations, while $\delta g_{mn}$ solves the
linearized equations. In general, evaluating the symplectic form away from
the solution space would be meaningless.

\subsection{Generalization to any Lagrangian theory}

\label{2.6}

The above strategy admits a natural generalization to any Lagrangian theory
\cite{CW},\cite{Zuckerman} (see also \cite{Wald}). For our purposes it will
be enough to assume that the Lagrangian $L=L(\phi^{A},\partial_{l}\phi^{A})$
(where the index $A$ numbers the fields) does not contain second- and
higher-order derivatives. Under these conditions, the \textit{Crnkovi\'{c}%
-Witten-Zuckerman} symplectic current is defined by%
\begin{equation}
J_{CWZ}^{l}=\delta\left( \frac{\partial L}{\partial\partial_{l}\phi^{A}}%
\right) \wedge\delta\phi^{A}\,.  \label{CWZ}
\end{equation}
The symplectic form is defined by (\ref{sform}), as before. Both properties
--- $\Sigma$-independence and gauge invariance --- are still true, provided
that the equations of motion are satisfied. The gravitational current (\ref%
{CW}) is obtained from the general formula (\ref{CWZ}), provided that one
drops the total second derivative term from the Einstein-Hilbert action,
which is equivalent to adding the Gibbons-Hawking boundary term \cite{GH}%
\footnote{%
It is convenient to use the explicit form of the Gibbons-Hawking Lagrangian $%
L_{\text{GH}}=\sqrt{-g}g^{ik}(\Gamma_{il}^{m}\Gamma_{km}^{l}-\Gamma_{ik}^{l}%
\Gamma_{lm}^{m})$ (see e.g.\ \cite{LLII}) when performing the computation
\cite{our}.}.

Another important invariance property of the symplectic current (\ref{CWZ})
is that its definition is independent of the choice of fundamental fields $%
\phi^{A}$. For example, in the case of pure gravity we may choose $\{\phi
^{A}\}=\{g_{mn}\}$ or $\{\phi^{A}\}=\{g^{mn}\}$, the resulting symplectic
currents being identical. The precise general statement is that the
symplectic current (\ref{CWZ}) is invariant under \textit{point
transformations} (i.e.\ transformations which do not involve derivatives of
the fields) $\phi\rightarrow\phi^{\prime}=\Phi(\phi)$. For a proof it is
convenient to rewrite the symplectic current as
\begin{equation}
J_{CWZ}^{l}=\delta I^{l},\qquad I^{l}=\frac{\partial L }{\partial\partial
_{l}\phi^{A}}\delta\phi^{A}\,,
\end{equation}
and show the invariance of $I^{l}$. Under a point transformation, $\delta
\phi^{A}$ transforms contravariantly in the $A$ index, i.e.\ it is
multiplied by the Jacobian:
\begin{equation}
\delta\phi^{A}\rightarrow\mathcal{J}^{A}{}_{B}\,\delta\phi^{B},\qquad
\mathcal{J}^{A}{}_{B}=\frac{\partial\Phi^{A}}{\partial\phi^{B}}\,.
\end{equation}
The derivatives $\partial_{l}\phi^{A}$ also transform contravariantly. Thus $%
\partial L/\partial\partial_{l}\phi^{A}$ will transform covariantly, and the
product $I^{l}$ is indeed invariant.

This invariance can be used to demonstrate the equivalence between the pure
gravity canonical symplectic form (\ref{can}) and the covariant expression (%
\ref{CW}). We just have to show that the gravitational symplectic form
evaluated using (\ref{CWZ}) reduces to (\ref{can}) if the ADM variables $%
N,N_{i},h_{ij}$ are used as a set of fields $\{\phi^{A}\}$ parametrizing the
metric. In a sense, this is obvious, because the $\Pi^{ij}$ given by (\ref%
{momenta}) were in fact defined by ADM \cite{ADM} exactly as the variation
w.r.t.\ $\dot{h}_{ij}$ of the gravitational Lagrangian, which written in
these variables takes the form%
\begin{equation}
L_{\text{ADM}}=N\sqrt{h}(K_{ij}K^{ij}-K^{2}+R^{(3)})\,.  \label{adml}
\end{equation}
In addition, time derivatives of $N$ and $N_{i}$ are absent from (\ref{adml}%
). Thus we see that the integrand of (\ref{can}) is nothing but $J^{t}$,
which is the only needed component of the symplectic current, since $\Sigma
=\{t=const\}$ in this parametrization\footnote{%
Strictly speaking, $L_{\text{ADM}}\neq L_{\text{GH}}$, because the $R^{(3)}$
term in (\ref{adml}) still contains second-order derivatives in \textit{%
spatial }directions. However, this difference does not affect the \textit{%
time} component of the symplectic current.}.

\subsection{Restriction of fixed fluxes}

\label{2.7}

As we stressed in the beginning, the on-shell quantization method
is designed to be used in the situations when all fluxes are
fixed. Actually more is true -- the method, at least in the
presented form, can be used \textit{only }in such situations. In
other words, the regimes of applicability of this method and of
flux quantization are mutually excluding.

To see this in a simple example, suppose that we use the on-shell
quantization method to quantize the Abelian gauge field on a 4-manifold of
nontrivial topology. From the action%
\begin{equation}
S=-\frac{1}{4}\int d^{4}x\sqrt{-g}F_{mn}F^{mn}
\end{equation}
we derive the symplectic current using (\ref{CWZ}):%
\begin{equation}
J^{l}=-\delta(\sqrt{-g}F^{lm})\wedge\delta A_{m\,.}
\end{equation}
We see that the gauge potential variation, $\delta A_{m}$, appears in this
expression. In order to compute the symplectic form, we have to be able to
choose a gauge in which $\delta A_{m}(x)$ is regular everywhere on the
Cauchy surface $\Sigma.$ This would be impossible unless the field strength
variation \ $\delta F=d(\delta A)$ has zero flux on all closed 2-cycles. It
is easy to see that this restriction is completely general and always
appears when there are gauge fields in the theory.

\begin{figure}[tb]
\begin{center}
\epsfig{figure=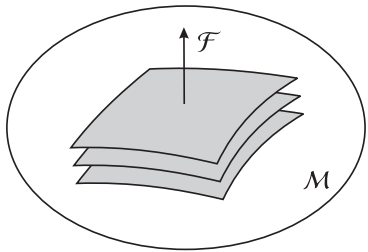,width=5cm}
\end{center}
\caption{An example of a 3-dimensional moduli space $\mathcal{M}$ foliated
by 2-dimensional symplectic sheets corresponding to a fixed value of a gauge
field flux $\mathcal{F}$. On-shell quantization can determine symplectic
structure on the sheets, but cannot be used to quantize the flux.}
\label{sheets}
\end{figure}

The restriction of fixed fluxes can be thought of as defining \emph{%
symplectic sheets} inside the moduli space, which thus becomes a Poisson
manifold \cite{Arnold}, with fluxes having trivial Poisson brackets with
everything else. Quantization inside each of these sheets is governed by a
corresponding symplectic form. Changing fluxes corresponds to continuously
changing the sheet. It is only when flux quantization is taken into account,
that this continuous variation becomes discrete (see Fig.\ \ref{sheets}).

\subsection{The role of dynamics}

\label{2.3} To get a nontrivial moduli space quantization, the symplectic
form computed by the on-shell quantization method should be non-degenerate
when restricted on $\mathcal{M}$. This implies that the canonical momenta
should be nontrivial functions on $\mathcal{M}$, much like in the chiral
boson example considered in Section \ref{2.2}, where we had $%
\pi=f^{\prime}(u)$. In the case of gravity this means that the class of
solutions cannot be static. Indeed, for static solutions the extrinsic
curvatures (\ref{curv}) are zero. Thus the canonical momenta (\ref{momenta})
will be identically zero on $\mathcal{M}$, and the restricted symplectic
form will vanish. This means that for static solutions, the parameters
characterizing the moduli space do not acquire nontrivial commutators among
themselves upon quantization.

The symplectic form will be usually non-degenerate if some dynamics is
present. For example, the \textquotedblleft Bubbling AdS" solutions
considered in Section \ref{3} below have nonzero angular momentum: they are
stationary but not static, and this will be enough to make the symplectic
form nondegenerate. The reason is that $N_{i}\neq0$ in (\ref{curv}) and the
extrinsic curvatures no longer vanish identically. In \cite{Mandal}, the
appearance of a nontrivial symplectic form on the moduli space was
conjectured to be related to supersymmetry. However, it is easy to see that
supersymmetry is neither necessary nor sufficient for that: the moduli
spaces of cylindrical \cite{Kuchar} or plane \cite{Mena} gravitational waves
have a nontrivial symplectic structure in pure gravity; on the other hand,
the symplectic form will vanish on any moduli space of static supersymmetric
solutions. Supersymmetry may be instrumental though to control corrections
to the on-shell quantization results (see Sections \ref{corr}) or to obtain
simplified symplectic current expressions (see remarks at the end of Section %
\ref{3.7.3}).

The above remarks are useful in order to understand the relation of our
method to the perhaps more familiar `moduli space approximation' \`{a} la
Manton \cite{Manton, Atiyah-Hitchin, Gibbons-Rudnick, Ferrell-Eardley},
where one usually starts with a moduli space describing multi-centered
static black hole or soliton solutions. This moduli space is typically
finite-dimensional, parametrized by the coordinates of the centers and
possibly some extra parameters. For example, in the multi-centered extremal
black hole case \cite{Ferrell-Eardley} the moduli space is $3n$-dimensional,
where $n$ is the number of black holes. Then one considers slow scattering
of these black holes (or solitons) and finds that it can be described by the
geodesics in a certain (calculable) metric%
\begin{equation}
ds^{2}=g_{ij}(q)dq^{i}dq^{j}\,.  \label{manton}
\end{equation}
Then one can consider quantization of this system of slowly moving solitons.
But notice that in the process of introducing slow motion we enlarged the
phase space by adjoining canonical momenta (corresponding to velocities).
So, in the black hole case, the phase space becomes $6n$-dimensional, twice
the dimensionality of the original moduli space. Also, in order to construct
these slowly moving black hole solutions (and derive metric (\ref{manton})),
one should go off-shell, i.e.\ out of the original moduli space. If one were
to recover the metric (\ref{manton}) using our method (which may or may not
be possible in practice), one would have to compute the symplectic form on
the extended $6n$-dimensional phase space. The symplectic form would be
degenerate if restricted to the original $3n$-dimensional moduli space
corresponding to the static configurations, in agreement with the above
discussion. This means in particular that black hole center coordinates are
not quantized by themselves.

\subsection{On-shell quantization vs effective action}

\label{versus}

In essence, the on-shell quantization method does nothing but quantizing
first-order fluctuations along the moduli space $\mathcal{M}$ (except that
it provides a particularly efficient way to do this). This is particularly
clear from the symplectic current expression (\ref{CWZ}), which involves
only first-order perturbations of the fields and, moreover, depends only on
the quadratic part of the Lagrangian as expanded around a given background.
To see this explicitly, let us parametrize the fields $\phi^{A}$ in (\ref%
{CWZ}) by their deviations $\psi^{A}$ from a given background $\phi_{0}\in%
\mathcal{M}$:%
\begin{equation}
\phi=\phi_{0}+\psi\,,
\end{equation}
at the same time expanding the Lagrangian of the theory in terms of $\psi$:%
\begin{equation}
L =\sum_{i=2}^{\infty} L ^{(i)},
\end{equation}
where $L ^{(i)}$ is degree $i$ in $\psi$. The part linear in $\psi$ is
absent, since we assume that $\phi_{0}$ is a solution. The coefficients of $%
L ^{(i)}$ may and will generically depend on $\phi_{0}$, but the precise
form of this dependence is irrelevant for our argument.

Now, as we showed in Section \ref{2.6}, the symplectic current is
independent of the field choice. In particular, we can use $\psi^{A}$ to
compute it. Thus we will have:
\begin{equation}
J_{CWZ}^{l}=\sum_{i=2}^{\infty}\delta\frac{\partial L ^{(i)}}{\partial
\partial_{l}\psi^{A}}\wedge\delta\psi^{A}.  \label{Jpsi}
\end{equation}
Since $L ^{(i)}$ is degree $i$ in $\psi,$ $\partial L ^{(i)}/\partial
\partial_{l}\psi$ will be degree ($i-1),$ and $\delta(\partial L
^{(i)}/\partial\partial_{l}\psi)$ will be degree $(i-2)$ in $\psi$ (and
linear in $\delta\psi$). So we see that once we evaluate (\ref{Jpsi}) on
shell by putting $\psi=0$, all the terms with $i\geq3$ will vanish, and thus
the symplectic current depends only on the quadratic part of the fluctuation
Lagrangian.

This remark makes it clear that any result obtained by the on-shell
quantization method can be in principle obtained by a more conventional
method based on computing the quadratic action, quantizing it, and
specializing to a particular subclass of perturbations corresponding to
deformations along $\mathcal{M}$. In practice, however, there may be
formidable difficulties in following the conventional path. The main
difficulty is that the full quadratic action will typically have a
complicated form, except around very special backgrounds. One could think
that perhaps one could consider a truncation of the quadratic action to the
fluctuation modes along $\mathcal{M}$, and that would have a chance to be
simple. The problem here is that such a truncation does not even have to
exist, since these modes can couple to other modes already on the quadratic
level.

These difficulties are best demonstrated on a concrete example of the LLM
family of solutions \cite{LLM}. On-shell quantization of this family is
considered in Section 3 below, and was also the subject of \cite{our}. In
this case the quadratic effective action is available around only one
representative of the family, namely $AdS_{5}\times S^{5}$ \cite{Shiraz,Kim}
(and can also be found around the plane wave, using the results of \cite{Metsaev}%
). In any other case it is highly unlikely that a tractable quadratic action
can be found. Moreover, even around $AdS_{5}\times S^{5}$ the quadratic
action couples the moduli space modes characterized by the BPS condition $%
J=\Delta$ to the opposite angular momentum modes $J=-\Delta$, as a simple
consequence of angular momentum conservation. This shows that on-shell
quantization is the only practical way to quantize in the general case.
However, around $AdS_{5}\times S^{5}$ both on-shell quantization and the
quadratic effective action method can be applied, and the results agree as
they should \cite{our}.

\subsection{Hilbert space. Semiclassical states. Hamiltonian}

\label{Hilbert}

Using the on-shell quantization method, we can compute the symplectic form
on the moduli space of solutions. This symplectic form encodes the Poisson
brackets between the functions defining the geometry (e.g.\ the shape of the
droplets in the LLM case). Quantizing the brackets, we find commutation
relations between these functions. We can then find a Hilbert space on which
these functions are defined as operators so that the commutation relations
are satisfied.

In principle, we will get a separate Hilbert space around each geometry from
the moduli space. Low-lying states in this Hilbert space will not allow any
semiclassical interpretation, they will be similar to states of a few field
quanta (e.g. gravitons) propagating in Minkowski space. However, in the
limit of large occupation numbers we can construct coherent states, which
can be interpreted as describing a neighboring classical geometry. In this
sense, neighboring geometries are contained in each other's Hilbert spaces.
This shows that all these Hilbert spaces will be isomorphic.

Let us now discuss the choice of a Hamiltonian operator on the Hilbert
space. First of all, when we are dealing with a theory of gravity, it is by
no means guaranteed that a preferred Hamiltonian will exist. For instance,
this seems to be the situation for the plane gravitational wave case
analyzed in \cite{Mena}. For the concept of energy to make sense, all
spacetimes in the considered moduli space must have a common asymptotic
infinity with a timelike Killing vector. In this case the Hamiltonian can be
defined by computing the classical energy of the considered solutions. For
the LLM case, such a computation has been done in \cite{LLM}.

When a Hamiltonian is available, the process of quantization can be taken
further by discussing the energy eigenstates. To avoid possible confusion we
would like to stress again: the Hamiltonian is an independent piece of
information which supplements the symplectic form computed by the on-shell
quantization method. There are cases when it can be defined in the classical
theory (and carried over to the quantum theory by the correspondence
principle), but a separate computation is necessary in order to do that.

\subsection{Corrections}

\label{corr}

In which sense does on-shell quantization approximate the true picture
attainable in the complete theory? And what are the corrections which should
be included if one is to go beyond this approximation? These are interesting
questions, which most likely have to be analyzed on a case-by-case basis.
Here we will limit ourselves to a few general remarks\footnote{%
In the context of minisuperspace approximation to pure gravity, some
discussion may also be found in \cite{Kuchar1}.}.

It will be convenient to phrase the discussion in the Hamiltonian language,
which may not be the most general situation (see Section \ref{Hilbert}), but
is sufficient for the applications we have in mind. The full Hamiltonian can
be schematically represented as
\begin{equation}
H=H_{0}(\phi_{\Vert})+H_{0}(\phi_{\perp})+H_{\text{int}}\,.
\end{equation}
Here $H_{0}(\phi_{\Vert})$ is the quadratic Hamiltonian for the fluctuation
modes along the moduli space $\mathcal{M}$ (around a given background); $%
H_{0}(\phi_{\perp})$ is the quadratic Hamiltonian for the modes
corresponding to fluctuations orthogonal to $\mathcal{M}$; $H_{\text{int}}$
is the interaction Hamiltonian.

On the quadratic level there is a complete decoupling between $\phi_{\Vert}$
and $\phi_{\perp}.$ It is only the $\phi_{\Vert}$ modes that we are
quantizing using the on-shell quantization method, while $\phi_{\perp}$ are
effectively frozen. The result of this quantization will be a Hilbert space $%
\mathcal{H}_{\Vert}$ and a spectrum of energy eigenstates\footnote{%
We do not discuss here the other aspect of the on-shell quantization, namely
the correspondence between coherent states and classical geometries
mentioned in Section \ref{Hilbert}.}.

It is more or less clear how this picture will have to be modified, if the
corrections are to be considered. First of all, we will have to introduce a
Hilbert space $\mathcal{H}_{\perp}$ for the orthogonal modes, with its own
free spectrum. The full Hilbert space is then the direct product $\mathcal{H}%
_{\Vert}\otimes\mathcal{H}_{\perp}$. On-shell quantization can be thought of
as turning off $H_{\text{int}}$ and putting all $\phi_{\perp}$ in their
vacuum state, so that the full state is a product%
\begin{equation}
|\phi_{\Vert}\rangle\otimes|0_{\perp}\rangle\,.  \label{product}
\end{equation}
Taking $H_{\text{int}}$ into account allows transitions, exciting
the orthogonal modes and inducing mixings between various states.
To be able to treat $H_{\text{int}}$ as a perturbation, a small
expansion parameter should be available. In this case the new
energy eigenstates will be small perturbations of the original
product states (\ref{product}), with slightly shifted energy
levels. In theories of gravity, the role of such an expansion
parameter can be played by $\ell\Lambda_{\text{Planck}}$, the
characteristic curvature radius $\ell$ of the background spacetime
measured in Planck
units. In the string theory context, this will correspond to the $%
\alpha^{\prime}$-expansion of the effective action.

In supersymmetric situations, the spectrum may be protected and energy
shifts should not occur. For instance, we know from AdS/CFT that this should
be the case for the LLM solutions. Indeed, the $\mathcal{N}=4$ SYM states
dual to these geometries are 1/2 BPS, and their energies cannot depend on a
continuous parameter. Notice that it would be wrong to conclude that the LLM
\textit{geometries} do not get modified once the $\alpha^{\prime}$%
-corrections are taken into account --- they will, except in the fully
supersymmetric cases of $AdS_{5}\times S^{5}$ and the plane wave. It is only
the energy eigenstates which should remain protected. It would be
interesting to show this directly from the gravity side.

The discussion of corrections becomes increasingly subtle when loop effects
are taken into account\footnote{%
Note that any discussion of such effects is ambiguous and ill-defined in
pure gravity, due to its non-renormalizability.}. In the AdS/CFT context,
the size of these effects is controlled by $1/N$. In this paper we mostly
discuss the $N=\infty$ limit, corresponding to the classical SUGRA. In
principle, it should be possible to consider $N\gg1$ case as a perturbation
over $N=\infty$. This would require inclusion of massive string states
needed for the UV completion of the theory. In the LLM case, we know that
this should lead to a reduction in the number of states with energies above $%
N$ (see Section \ref{3.3}). It would be extremely interesting to understand
how such a reduction can be achieved in a perturbative treatment, and to
reproduce it from the gravity side.

\subsection{Summary of on-shell quantization method}

\label{2.8}

To summarize, we have the following recipe for quantizing a moduli space of
solutions of any supergravity theory. First of all, we have to find a
general expression for the symplectic current of the theory. This is
computed from the supergravity Lagrangian by the general formula (\ref{CWZ})
and will contain the Crnkovi\'{c}-Witten gravitational symplectic current (%
\ref{CW}), as well as additional terms for the other fields of the theory.

Second, we have to evaluate the symplectic current on the moduli space of
solutions. This is done by expressing the variations of all the fields in
the symplectic current via the variations of the arbitrary functions
describing the moduli space.

Finally, we have to integrate the symplectic current on a Cauchy surface $%
\Sigma$ to obtain the symplectic form, which will be a closed 2-form on the
moduli space. This symplectic form can then be quantized in the standard way.

\section{Quantization of ``Bubbling AdS"}

\label{3}

\subsection{Supergravity solutions}

\label{3.1} Having set up the general framework in the previous section, we
will now apply the on-shell quantization method to quantize the
\textquotedblleft Bubbling AdS" family of supergravity solutions found by
LLM \cite{LLM}. This family includes all regular 1/2 BPS solutions of Type
IIB SUGRA with $SO(4)\times SO(4)\times\mathbb{R}$ symmetry. These solutions
have constant dilaton and axion and vanishing 3-form. The metric has the
form:%
\begin{equation}
ds^{2}=-h^{-2}(dt+V_{i}dx^{i})^{2}+h^{2}(dy^{2}+dx^{i}dx^{i})+ye^{G}d%
\Omega_{3}^{2}+ye^{-G}d\tilde{\Omega}_{3}^{2}\,,
\end{equation}
where $i=1,2$ , $d\Omega_{3}^{2}$ and $d\tilde{\Omega}_{3}^{2}$ are the
metrics on two unit 3-spheres $S^{3},\tilde{S}^{3}.$ The functions $h$ and $%
G $ are determined in terms of a single function $z(x_{1},x_{2},y),$ $y>0$:%
\begin{equation}
h^{-2}=\frac{y}{\sqrt{1/4-z^{2}}}\,,\qquad e^{2G}=\frac{1/2+z}{1/2-z}\,,
\label{hG}
\end{equation}
The $z$ and $V_{i}$ are in turn fixed in terms of one function $%
Z(x_{1},x_{2}),$ which can only take the values $\pm\frac{1}{2},$ and is
also the boundary value of $z$ on the $y=0$ plane. Namely, we have\footnote{%
We use here the standard notation for two-dimensional convolution: $f\ast
g(x)=\int f(x-x^{\prime})g(x^{\prime})d^{2}x^{\prime}$.}:%
\begin{align}
z & =\frac{1}{\pi}\frac{y^{2}}{(x^{2}+y^{2})^{2}}\ast Z\,,  \notag \\
V_{i} & =\frac{\varepsilon^{ij}}{\pi}\frac{x_{j}}{(x^{2}+y^{2})^{2}}\ast Z\,.
\label{V}
\end{align}
Apart from the metric, only the 5-form is turned on:%
\begin{align}
F_{5} & =F\wedge d\Omega_{3}+\tilde{F}\wedge d\tilde{\Omega}_{3}\,,  \notag
\\
F & =dB,\qquad\tilde{F}=d\tilde{B},\qquad
\end{align}
$d\Omega_{3},d\tilde{\Omega}_{3}$ being the volume forms of the spheres. The
one-forms $B,\tilde{B}$ are defined up to a gauge transformation. A
convenient choice is the axial gauge $B_{y}=\tilde{B}_{y}=0,$ the remaining
components being then given by
\begin{align}
B_{t} & =-\frac{1}{4}y^{2}e^{2G}, & \tilde{B}_{t} & =-\frac{1}{4}%
y^{2}e^{-2G},  \label{bt} \\[0.06in]
B_{i} & =-\frac{y^{2}V_{i}}{4\left( {\frac{1}{2}}-z\right) }-\frac{U_{i}}{4}-%
\frac{x_{1}}{4}\delta_{i,2}\,, & \tilde{B}_{i} & =-\frac{y^{2}V_{i}}{4\left(
{\frac{1}{2}}+z\right) }-\frac{U_{i}}{4}+\frac{x_{1}}{4}\delta_{i,2}\,
\label{axial}
\end{align}
(see \cite{our}), where%
\begin{equation}
U_{i}\equiv\frac{\varepsilon^{ij}}{\pi}\frac{x_{j}}{x^{2}+y^{2}}\ast Z\,.
\label{U}
\end{equation}
The following linear relations, evident from (\ref{V}) and (\ref{U}), will
play an important role below:%
\begin{align}
\partial_{i}z & =-y\,\varepsilon^{ij}\partial_{y}V_{j}\,,  \notag \\
\partial_{\lbrack i}V_{j]} & \equiv\frac{1}{2}(\partial_{i}V_{j}-%
\partial_{j}V_{i})=\frac{1}{2y}\varepsilon_{ij}\partial_{y}z\,,
\label{linear} \\
\partial_{y}U_{i} & =-2yV_{i}\,.  \notag
\end{align}

\subsection{Moduli space}

\label{3.2} The moduli space is parametrized by collections of droplets of
arbitrary shape in the $x_{1},x_{2}$ plane, defined by the condition that $%
Z=-1/2$ on the union of all droplets (which we denote $\mathcal{D})$ and $%
Z=1/2$ in the remaining part of the plane, $\mathcal{D}^{c}$. We will assume
that the boundaries of all droplets in $\mathcal{D}$ are smooth curves, so
that the corresponding solution has a smooth geometry \cite{LLM}. We would
like to discuss the quantization of this moduli space. As discussed in
Section \ref{2}, there are two aspects to the moduli space quantization.
First we must detect all nontrivial fluxes. These can be quantized using an
appropriate Dirac quantization condition. On the other hand, deformations
within the moduli space corresponding to keeping all the fluxes fixed should
be quantized using the on-shell quantization method.

\begin{figure}[tb]
\centering {\ \includegraphics[width=5cm]{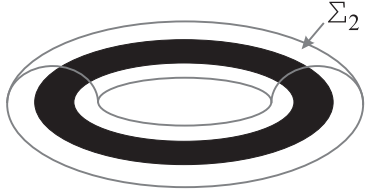} }
\caption{The surface $\Sigma_{2}$ caps the droplet in the $x_{1},x_{2}$
plane by extending into $y>0$. Fibering the $\tilde{S}^{3}$ over $\Sigma$,
we get a closed 5-manifold supporting an $F_{5}$ flux proportional to the
area of the droplet \protect\cite{LLM}.}
\label{capping}
\end{figure}

The topology of the \textquotedblleft Bubbling AdS" spacetimes and the
corresponding fluxes have been already determined in \cite{LLM}. It turns
out that for every black or white region in the droplet plane there is a
non-contractible 5-cycle supporting an $F_{5}$ flux proportional to the area
of the corresponding region. For example, for each droplet (i.e.\ a black
region) the corresponding 5-manifold is constructed by fibering the sphere $%
\tilde{S}^{3}$ over a surface $\Sigma_{2}$ capping the droplet as shown in
Fig.~\ref{capping}. This 5-manifold is nonsingular, since the $\tilde{S}^{3}$
shrinks to zero size on the $y=0$ plane outside of the droplets in an
appropriate way. Analogously, the 5-manifolds corresponding to the white
regions are constructed by fibering the $S^{3}$ over surfaces capping these
white `holes' inside the droplets. Using flux quantization, one shows that
the area of each black or white region must be quantized \cite{LLM}\footnote{%
The relation $\kappa_{10}=8\pi^{7/2}\ell_{P}^{4}$ is useful in comparing
some of our equations to \cite{LLM}.}:
\begin{equation}
A_{i}=\frac{\kappa_{10}}{2\pi^{3/2}}n_{i},\qquad n_{i}\in\mathbb{N}.
\label{area}
\end{equation}
These integers can be thought of as the numbers of giant gravitons wrapping
the $S^3$ and the $\tilde{S}^3$. The sum of $n_{i}$'s corresponding to the
black regions should be equal $N,$ the total number of D3 branes making up
the configuration.

We see that there exists an infinite-dimensional class of deformations
within the moduli space keeping all the fluxes fixed: these are precisely
the deformations of the boundaries of the droplets which keep fixed the
areas of all black and white regions. It is these deformations that we will
quantize using the on-shell quantization method.

\subsection{Dual description in terms of free fermions}

\label{3.3} The \textquotedblleft Bubbling AdS" solutions corresponding to a
collection of finite-size droplets are asymptotically $AdS_{5}\times S^{5}$.
The AdS/CFT correspondence \cite{AdSCFT} can be used to relate these Type
IIB supergravity solutions to states of the $\mathcal{N}=4$ supersymmetric
Yang-Mills theory on $S^{3}\times\mathbb{R}$. Since the geometries are 1/2
BPS, the dual operators on the Yang-Mills side are chiral primaries with
conformal weight equal to their $U(1)$ R-charge: $\Delta=J$. As is well
known \cite{Corley,Berenstein}, this sector of $\mathcal{N}=4$ super
Yang-Mills admits a very simple description as a system of $N$
non-relativistic free fermions moving in a harmonic oscillator potential. In
the large $N$ limit (which is what we are mostly concerned with in this
paper) the states of the many-fermion system are well described as droplets
in the one-particle phase space. In fact, these droplets are the same
droplets which characterize the gravity solutions. For example, it was
checked in \cite{LLM} that the excitation energy of the \textquotedblleft
Bubbling AdS" solutions over $AdS_{5}\times S^{5}$ is equal to%
\begin{equation}
\Delta=J=\frac{1}{4\pi\hbar^{2}}\left[ \int_{\mathcal{D}}d^{2}x\,x^{2}-\frac{%
1}{2\pi}\left( \int_{\mathcal{D}}d^{2}x\right) ^{2}\right] \,,
\end{equation}
which is precisely the excitation energy of the fermionic state described by
the droplet $\mathcal{D}$ over the ground state described by the circular
droplet of the same area centered at the origin. The one-dimensional Planck
constant $\hbar$ here can be fixed by comparing the area quantization
condition (\ref{area}) valid on the supergravity side with the semiclassical
phase space quantization condition $A=2\pi\hbar n$, which should be true on
the fermion side. This gives \cite{LLM}:%
\begin{equation}
\hbar=\frac{\kappa_{10}}{4\pi^{5/2}}\,.  \label{hbar}
\end{equation}

In this paper we would like to compare the symplectic structures on the
droplet space arising from the supergravity and the fermion side. On the
supergravity side we will use the on-shell quantization method described in
Section \ref{2}. On the fermion side the corresponding symplectic form can
be found by using the so-called hydrodynamic approach, commonly used to
describe edge states in Quantum Hall systems (see e.g. \cite{Wen}). The idea
of this method is to identify the one-fermion description of the classical
dynamics of the droplets with the collective one. In the one-fermion
picture, the droplet motion is described by solving the individual Hamilton
equations\footnote{%
This normalization of the Hamiltonian has to be chosen so that the energy
levels become integer-spaced and can be in one-to-one correspondence with
the integer dimensions of field theory chiral primaries.}%
\begin{equation}
\dot{p}=-\partial_{q}H,\quad\dot{q}=\partial_{p}H,\quad H=\frac{p^{2}+q^{2}}{%
2\hbar}\,.  \label{one}
\end{equation}
for the fermions localized close to the droplet boundary $\partial\mathcal{D}
$. It is convenient to first look at the subclass of the droplets whose
boundary curve can be described in polar coordinates as $r=r(\phi)$ for a
single-valued function $r(\phi)$. For these droplets (\ref{one}) implies
that $r(\phi)$ evolves in time according to the classical chiral boson
equation%
\begin{equation}
\dot{r}=\hbar^{-1}r^{\prime}.  \label{chiral}
\end{equation}
(A cautionary remark: as we will see below, it would be wrong to conclude
from this that the in quantum theory $r(\phi)$ satisfies the chiral boson
commutation relations.) In the collective pictures, one would like to
recover the same equation (\ref{chiral}) as a Hamilton equation of the form%
\begin{equation}
\dot{r}=\{r,H_{\text{tot}}\}
\end{equation}
where $H_{\text{tot}}$ is the total energy of the droplet state, given by
the integral of the one-particle Hamiltonian:%
\begin{align}
H_{\text{tot}} & =\int_{\mathcal{D}}\frac{dp\,dq}{2\pi\hbar}\frac{p^{2}+q^{2}%
}{2\hbar}  \notag \\
& =\,\frac{1}{16\pi\hbar^{2}}\oint_{\partial\mathcal{D}}d\phi\,r^{4}(\phi),
\label{ham}
\end{align}
while the Poisson bracket $\{,\}$ is to be determined from consistency of
the two pictures. As we showed in \cite{our} (see also \cite{Poly}, and
especially \cite{Dhar}, Eq.\ (4.3)), the appropriate Poisson bracket which
generates the correct equation is%
\begin{equation}
\{r^{2}(\phi),r^{2}(\tilde{\phi})\}=8\pi\hbar\,\delta^{\prime}(\phi -\tilde{%
\phi}).  \label{fermPB}
\end{equation}
The symplectic form corresponding to these Poisson brackets can be written as%
\footnote{%
The Poisson brackets are encoded by the symplectic form in the following
schematic way: $\{q_{i},q_{j}\}_{P.B.}=A_{ij}$ corresponds to $\omega=\frac{1%
}{2}A_{ij}^{-1}dq_{i}\wedge dq_{j}.$To apply this standard rule in our
situation, note that the inverse of the kernel $\delta^{\prime}(\phi-\tilde{%
\phi})$ is $\frac{1}{2}$Sign$(\phi-\tilde{\phi}).$}:%
\begin{equation}
\omega_{\text{ferm}}=\frac{1}{32\pi\hbar}\oint\oint d\phi\,d\tilde{\phi }\,%
\text{Sign}(\phi-\tilde{\phi})\,\delta\left[ r^{2}(\phi)\right] \wedge\delta%
\left[ r^{2}(\tilde{\phi})\right] \,.  \label{fermform}
\end{equation}
The total area of the droplet is proportional to the total number of
fermions and must be kept fixed. This means that the variations in (\ref%
{fermform}) must satisfy the constraint%
\begin{equation}
\oint d\phi\,\,\delta\left[ r^{2}(\phi)\right] =0\,.  \label{constr}
\end{equation}
Under this constraint, the symplectic form (\ref{fermform}) is well defined
as written, in spite of the fact that the kernel Sign$(\phi-\tilde{\phi})$
is not periodic. In fact we can add to the kernel any function of the form $%
g_{1}(\phi)+g_{2}(\tilde{\phi})$ without changing the answer. If desired,
one can consider adding $(\tilde{\phi}-\phi)/\pi$, which will make the
kernel explicitly periodic.

It is interesting to note that although the symplectic form (\ref{fermform})
is easiest to derive for the particular case of the harmonic oscillator
one-fermion Hamiltonian, it is in fact completely general and will describe
the motion of droplets of noninteracting fermions described by an arbitrary
one-particle Hamiltonian $H(p,q)$ \cite{Poly},\cite{our},\cite{Dhar}.

Let us rewrite (\ref{fermform}) in a slightly different form, introducing
instead of $\phi$ a natural parameter $s$ measuring arc length along $%
\partial\mathcal{D}$. From elementary geometry we have
\begin{equation}
\frac{ds}{r(\phi)d\phi}=\frac{\delta r}{\delta\gamma_{\perp}}\,,
\label{elem}
\end{equation}
where by $\delta\gamma_{\perp}$ we denote the variation of $\partial
\mathcal{D}$ in the outside normal direction. Thus (\ref{fermform}) can be
equivalently written as
\begin{equation}
\omega_{\text{ferm}}=\frac{1}{8\pi\hbar}\oint\oint ds\,d\tilde{s}\,\text{Sign%
}(s-\tilde{s})\,\delta\gamma_{\perp}(s)\wedge\delta\gamma_{\perp }(\tilde{s}%
)\,.  \label{general}
\end{equation}
The constraint (\ref{constr}) takes the form
\begin{equation}
\oint ds\,\delta\gamma_{\perp}(s)=0\,.  \label{constr1}
\end{equation}
The advantage of expression (\ref{general}) is that it is completely general
--- it makes no reference to polar coordinates and can be also used for
droplets with multiple-valued $r(\phi)$. It is this expression that we will
have to reproduce from the supergravity side.

The Poisson bracket following from (\ref{general}),%
\begin{equation}
\{\delta\gamma_{\perp}(s),\delta\gamma_{\perp}(\tilde{s})\}=2\pi\hbar
\,\delta^{\prime}(s-\tilde{s})\,,  \label{bg1}
\end{equation}
is equivalent to the bracket derived by Dhar \cite{Dhar}, Eq.\ (4.2), using
a reparametrization-invariant description of fermion droplet boundaries. In
Appendix \ref{A.0} we show explicitly that (\ref{bg1}) generates correct
equations of motion of droplet boundary in the general case.

A few words should be said about the symplectic structure in the situation
when several droplets are present, or when a single droplet has several
boundary components. Assume that $\partial\mathcal{D}$ has $B$ connected
components described by closed curves $\gamma^{(b)}(s)$. In this case one
has to introduce a separate field $\,\delta\gamma_{\perp}^{(b)}(s)$ for each
boundary. The total symplectic form is given by the sum of the forms (\ref%
{general}) computed for each boundary component:

\begin{equation}
\omega_{\text{ferm}}=\frac{1}{8\pi\hbar}\sum_{b=1}^{B}\oint\oint_{\gamma
^{(b)}}ds\,d\tilde{s}\,\text{Sign}(s-\tilde{s})\,\delta\gamma_{%
\perp}^{(b)}(s)\wedge\delta\gamma_{\perp}^{(b)}(\tilde{s})\,.
\label{genform}
\end{equation}
Each $\delta\gamma_{\perp}^{(b)}(s)$ has to satisfy its own constraint%
\begin{equation}
\oint_{\gamma^{(b)}}\delta\gamma_{\perp}^{(b)}(s)\,ds=0\,.  \label{own}
\end{equation}
This means that, semiclassically, different droplet boundaries are
completely decoupled, and it is impossible for a fermion to move from one
boundary to another, even though such a transition would keep the total area
of the droplet fixed. Technically, condition (\ref{own}) specifies
symplectic sheets in the moduli space of fermion droplets.

The symplectic form (\ref{genform}) on variations satisfying (\ref{own}) is
all we need for future comparison with the supergravity side, since as we
discussed in the previous subsection, only variations keeping both black and
white areas fixed can be quantized by the on-shell quantization method, due
to the fixed flux restriction.

Knowing the Poisson brackets, it is trivial to quantize the system. We will
do this around the circular droplet, the general case being identical. We
start by expanding $r^{2}(\phi)$ in Fourier series:%
\begin{equation}
r^{2}(\phi)=\sum\alpha_{n}e^{in\phi},\qquad\alpha_{-n}=\alpha_{n}^{\ast }.
\label{fexp}
\end{equation}
The zero mode is fixed in terms of the droplet area:%
\begin{equation}
\alpha_{0}=\frac{A}{\pi}.
\end{equation}
The $n\neq0$ modes correspond to the area-preserving deformations.
Substituting (\ref{fexp}) in (\ref{fermform}), we get Poisson brackets in
terms of the corresponding Fourier coefficients:%
\begin{equation}
\{\alpha_{n},\alpha_{m}\}=-4\hbar in\delta_{n+m}\,,  \label{torepr}
\end{equation}
which upon quantization become commutators:%
\begin{equation}
\lbrack\alpha_{n},\alpha_{m}]=4\hbar^{2}n\delta_{n+m}\,,
\end{equation}
The Hilbert space of the theory can be constructed as a bosonic Fock space
with $\alpha_{n},$ $n>0$ $(n<0)$ as annihilation (creation) operators. The
Hamiltonian can be expressed in terms of these operators as
\begin{equation}
H=\frac{1}{4\hbar^{2}}\sum_{n>0}\alpha_{-n}\alpha_{n}+\text{Const.}
\end{equation}
It is interesting to compare the partition function of this theory%
\begin{equation}
Z(\beta)=\prod_{n=1}^{\infty}\frac{1}{1-e^{-\beta n}}\,.
\end{equation}
with the exact partition function for a system of $N<\infty$ fermions in the
harmonic oscillator potential:%
\begin{equation}
Z_{N}(\beta)=\prod_{n=1}^{N}\frac{1}{1-e^{-\beta n}}\,.  \label{zexact}
\end{equation}
Writing the partition functions in the form
\begin{equation}
Z(\beta)=\sum_{E_{i}}g(E_{i})e^{-\beta E_{i}}\,,
\end{equation}
we can read off the degeneracy $g(E_{i})$ of the energy state $E_{i}$. We
see that the infinite $N$ computation correctly predicts the finite $N$
energy levels $E_{i}=1,2,3,\ldots$ However, it reproduces their degeneracies
only up to energies $E_{i}\leq N,$ overestimating the degeneracy of higher
energy levels\footnote{%
This was first pointed out to us by S. Minwalla.}.

\subsection{Symplectic form of Type IIB SUGRA}

\label{3.4} The LLM solutions satisfy the equations of motion that can be
derived from the following consistent truncation of the Type IIB SUGRA action%
\footnote{%
The notation $|i_{1}\ldots i_{n}|$ means that the indices have to be
ordered: $i_{1}<\ldots<i_{n}.$ Thus we have $F=F_{|i_{1}\ldots
i_{5}|}dx^{i_{1}}\wedge\ldots\wedge dx^{i_{5}}=\frac{1}{5!}F_{i_{1}\ldots
i_{5}}dx^{i_{1}}\wedge\ldots\wedge dx^{i_{5}}.$ The same ordering is assumed
in the summation.}:
\begin{equation}
S=\frac{1}{2\kappa_{10}^{2}}\int d^{10}x\sqrt{-g}\left( R-4F_{|i_{1}\ldots
i_{5}|}F^{|i_{1}\ldots i_{5}|}\right) ,  \label{IIB}
\end{equation}
We can proceed using this action if we impose the selfduality constraint $%
F_{5}=\ast F_{5}$ on the solutions. The presence of this constraint does not
modify the underlying symplectic form of the theory, which can be computed
from the action (\ref{IIB}).

Using the general formulas from Section \ref{2}, the symplectic form will be
equal to\footnote{%
The symplectic form and currents appearing in this section were already
derived in \cite{our}.}%
\begin{equation}
\omega=\frac{1}{2\kappa_{10}^{2}}\int d\Sigma_{l}(J_{G}^{l}+J_{F}^{l}),
\label{sformIIB}
\end{equation}
where $J_{G}^{l}$ and $J_{F}^{l}$ are symplectic currents constructed from
the gravity and 5-form parts of the Langrangian using (\ref{CWZ}). In fact $%
J_{G}^{l}$ is nothing but the Crnkovi\'{c}-Witten current (\ref{CW}). To
find the 5-form current$,$ we take the potentials $A_{|k_{1}\ldots k_{4}|}$ $%
(F_{5}=dA)$ as our basic fields. Applying (\ref{CWZ}), we get immediately
\begin{equation}
J_{F}^{l}=-8\,\delta(\sqrt{-g}F^{l|k_{1}\ldots k_{4}|})\,\wedge\delta
A_{|k_{1}\ldots k_{4}|}.  \label{jfexp}
\end{equation}
Now that we computed the symplectic current, we can simplify it using the
selfduality constraint, which can be written as
\begin{equation}
\sqrt{-g}F^{l_{1}\ldots l_{5}}=\varepsilon^{l_{1}\ldots l_{5}|m_{1}\ldots
m_{5}|}F_{|m_{1}\ldots m_{5}|},
\end{equation}
where $\varepsilon^{\ldots}$ is the flat 10-dimensional epsilon symbol. Thus
we have%
\begin{equation}
J_{F}^{l}=-8\,\varepsilon^{l|k_{1}\ldots k_{4}||m_{1}\ldots m_{5}|}\delta
F_{|m_{1}\ldots m_{5}|}\wedge\delta A_{|k_{1}\ldots k_{4}|}.
\label{jfexpsimp}
\end{equation}

\subsection{Regularity condition. Boundary term}

\label{3.5} To evaluate the symplectic form, we will use $\Sigma =\{t=const\}
$ in (\ref{sform}). This is a natural choice, in view of the
time-independence of the solutions. With this choice, $J^{t}$ is the only
needed component of the symplectic current. Because of the $SO(4)\times SO(4)
$ symmetry, integration over the 3-spheres is trivial, and we will have%
\begin{equation}
\omega =\frac{(2\pi ^{2})^{2}}{2\kappa _{10}^{2}}\int
dy\,d^{2}x(J_{G}^{t}+J_{F}^{t})\,.  \label{integral}
\end{equation}%
There is however a subtlety to be discussed, before we start evaluating this
integral. Namely, we have to make sure that the field perturbations we are
using are \textit{regular}. More precisely, this condition means that $%
\delta g_{mn}$ and $\delta A_{4}$ should be regular in a local coordinate
system in which the background metric $g_{mn}$ is regular (such a coordinate
system exists because all LLM geometries are regular). If this condition is
violated, we may get a wrong result for the symplectic form. In fact, as we
will explain below, this was precisely the origin of the factor 1/2
discrepancy encountered in \cite{our}. The source of trouble is $\delta
A_{4},$ which has the form%
\begin{equation}
\delta A_{4}=\delta B\wedge d\Omega +\delta \tilde{B}\wedge d\tilde{\Omega}\,
\label{dA}
\end{equation}%
and does not satisfy the regularity condition on the $y=0$ plane if the
axial gauge expressions (\ref{axial}) are used. In fact, $\delta B$ and $%
\delta \tilde{B}$ obtained by perturbing (\ref{axial}) tend to a finite
nonzero limit as $y\rightarrow 0$. On the other hand, the spheres $S^{3}(%
\tilde{S}^{3})$, whose volume forms appear in (\ref{dA}), shrink to zero
size inside (outside) the droplets. This means that the four-form (\ref{dA})
is singular as $y\rightarrow 0$.\

To get the right answer for the symplectic form, we must use in (\ref%
{jfexpsimp}) a regular gauge field perturbation $\delta A_{4}^{\text{reg}}$.
The existence of such a regular perturbation and its properties are discussed in
detail in Appendix \ref{regul}. The result of this analysis is that $\delta
A_{4}^{\text{reg}}$ can be chosen to have the form
\begin{equation}
\delta A_{4}^{\text{reg}}=\delta B^{\text{reg}}\wedge d\Omega +\delta \tilde{%
B}^{\text{reg}}\wedge d\tilde{\Omega}\,,  \label{dAreg}
\end{equation}
where $\delta B^{\text{reg}},\delta \tilde{B}^{\text{reg}}$ are regular
functions at $y\geq 0$, which have zero $y=0$ limits on $\mathcal{D}$($%
\mathcal{D}^{c}$) as a consequence of the regularity of $\delta A_{4}^{\text{%
reg}}$:%
\begin{eqnarray}
\delta B^{\text{reg}}|_{y=0} &=&0\quad \text{on }\mathcal{D},  \notag \\
\delta \tilde{B}^{\text{reg}}|_{y=0} &=&0\quad \text{on }\mathcal{D}^{c}.
\label{reg}
\end{eqnarray}
(the precise regularity condition is somewhat stronger, see appendix \ref{regul}).

Since $\delta A_{4}^{\text{reg}}$ must be related to $\delta A_{4}$ by a
gauge transformation, we must have%
\begin{equation}
\delta B^{\text{reg}}=\delta B-d\lambda \,,\qquad \delta \tilde{B}^{\text{reg%
}}=\delta \tilde{B}-d\tilde{\lambda}\,  \label{gtr}
\end{equation}%
for some regular functions $\lambda ,\tilde{\lambda}$.

Now we are ready to find the 5-form part of the symplectic form. First of
all, substituting (\ref{dAreg}) into (\ref{jfexpsimp}), we get

\begin{equation}
J_{F}^{t}=4\varepsilon ^{abc}(\,\delta B_{a}^{\text{reg}}\wedge \delta
\tilde{F}_{bc}-\delta \tilde{B}_{a}^{\text{reg}}\wedge \delta F_{bc}\,)\,\,.
\end{equation}%
Using (\ref{gtr}), this can be expressed as
\begin{align}
J_{F}^{t}& =4\varepsilon ^{abc}(\,\delta B_{a}\wedge \delta \tilde{F}%
_{bc}-\delta \tilde{B}_{a}\wedge \delta F_{bc}\,)\,-4\varepsilon
^{abc}(\,\partial _{a}\lambda \wedge \delta \tilde{F}_{bc}-\partial _{a}%
\tilde{\lambda}\wedge \delta F_{bc}\,)\,  \notag \\
& =8\varepsilon ^{ij}(\partial _{y}\delta B_{i}\,\wedge \delta \tilde{B}%
_{j}-\delta B_{i}\,\wedge \partial _{y}\delta \tilde{B}_{j})-4\varepsilon
^{abc}\partial _{a}(\,\lambda \wedge \delta \tilde{F}_{bc}-\tilde{\lambda}%
\wedge \delta F_{bc}\,)\,\,,
\end{align}%
where we simplified the first term using the axial gauge condition $B_{y}=%
\tilde{B}_{y}=0$ and used the closedness of $F$ and $\tilde{F}$ to rewrite
the second term as a total derivative. Contribution of this term to the
integral in (\ref{integral}) will be given by a boundary term at $y=0$. More
precisely, we will have%
\begin{equation}
\int dy\,d^{2}x\,J_{F}^{t}\,=\int dy\,d^{2}x\,J_{F}^{\text{bulk}%
}+\int_{y=0}d^{2}x\,J_{F}^{\text{bdry}}\,,  \label{corrected}
\end{equation}%
where%
\begin{align}
J_{F}^{\text{bulk}}& =8\,\varepsilon ^{ij}(\partial _{y}\delta B_{i}\,\wedge
\delta \tilde{B}_{j}-\delta B_{i}\,\wedge \partial _{y}\delta \tilde{B}%
_{j})\,,  \notag \\
J_{F}^{\text{bdry}}& =8\,(\lambda \wedge \delta \tilde{F}_{12}-\tilde{\lambda%
}\wedge \delta F_{12}\,)\,\,.  \label{currents}
\end{align}%
If we had used the singular perturbation $\delta A_{4}$ in (\ref{jfexpsimp}%
), we would have missed the boundary term in (\ref{corrected}). This is what
caused the factor 1/2 discrepancy encountered in Section 6 of \cite{our}.

Eq. (\ref{corrected}) is the final general expression for the 5-form part of
the symplectic form. We see that to evaluate it, we only need to know the
functions $\lambda $ ($\tilde{\lambda}$) at $y=0$. Their values on $\mathcal{%
D}$ ($\mathcal{D}^{c}$) can be determined from (\ref{reg}). Using (\ref{gtr}%
), we see that we must solve the equations%
\begin{align}
\delta B_{i}& =\partial _{i}\lambda \quad \text{on }\mathcal{D},  \notag \\
\delta \tilde{B}_{i}& =\partial _{i}\tilde{\lambda}\quad \text{on }\mathcal{D%
}^{c}.\label{tosolve}
\end{align}%
In the complementary regions $\lambda $ ($\tilde{\lambda}$) can be defined
arbitrarily, since the field strengths multiplying these functions in $%
J_{F}^{\text{bdry}}$ will vanish there as a consequence of (\ref{tosolve}).
This is of course not surprising, since making $\lambda $ or $\tilde{\lambda}
$ nonzero in the complementary regions corresponds to a perfectly regular
gauge transformation.

Let us repeat the logic of the preceding discussion. We observed that the
axial-gauge perturbation $\delta A_{4}$ is singular at $y=0.$ We must
therefore gauge-transform $\delta A_{4}$ into a regular perturbation $\delta
A_{4}^{\text{reg}}$. Such a regular form must exist, since all the fluxes
are kept fixed. In principle, it is $\delta A_{4}^{\text{reg}}$ which should
be used to compute the symplectic form. However, in practice it would be
convenient to be able to use the explicit axial gauge expressions. Eq.\ (\ref%
{corrected}) shows that we \textit{can} keep using $\delta A_{4}$ to compute
the symplectic current, but we have to pay a small price of adding a term to
the symplectic form localized on the lower-dimensional submanifold where $%
\delta A_{4}$ becomes singular, i.e.\ at $y=0$. We will refer to this term
as the \textit{boundary term}, although of course it should be remembered
that $y=0$ is only a coordinate boundary and the spacetime is completely
regular there.

It remains to discuss the regularity of $\delta g_{mn}$. This perturbation
is manifestly regular everywhere at $y=0$ except maybe at $\partial\mathcal{D%
}$. Thus we see that the situation is better than with $\delta A_{4},$
already because the singularity occurs on a smaller set. In fact, one can
show that the remaining singularity on $\partial\mathcal{D}$ is harmless,
and does not give rise to any additional terms in the symplectic form. The
argument, which we will not present here in detail, proceeds by constructing
a linear gauge transformation which completely removes the singularity in $%
\delta g_{mn},$ and by showing that this gauge transformation has no effect
on the symplectic form. Thus the gravitational part of the symplectic form
can be computed by using the naive expressions for $\delta g_{mn}$ in the
Crnkovi\'{c}-Witten symplectic current (\ref{CW}).

\subsection{Wavy line approximation}

\label{3.6} We will now use our method to compute the symplectic form around
the plane wave background, described by the droplet filling the lower
half-plane: $\mathcal{D}=\{x_{2}<0\}$. The fluctuations around this
background correspond to deforming $\partial\mathcal{D}$ from $x_{2}=0$ into
$x_{2}=\varepsilon(x_{1}).$ The symplectic form will be an antisymmetric
bilinear functional of $\varepsilon(x)$. The most general form of such a
functional is%
\begin{equation}
\omega=\int\int dx\,d\tilde{x}\,K(x,\tilde{x})\,\varepsilon(x)\wedge
\varepsilon(\tilde{x})\,,\quad K(\tilde{x},x)=-K(x,\tilde{x})\,.
\end{equation}
Since $x_{1}\rightarrow x_{1}+const$ is a symmetry of the plane wave
background, the kernel $K(x,\tilde{x})$ will be translationally invariant: $%
K(x,\tilde{x})=K(x-\tilde{x})$. This suggests that we should work in the
Fourier-transformed basis, in which $\omega$ will have a diagonal form%
\footnote{%
To avoid proliferation of tildes, we denote Fourier-transformed
functions by
the same symbol with $p$ as an argument.}:%
\begin{align}
\omega= & \int\frac{dp}{2\pi}\,K(p)\,\varepsilon(p)\wedge\varepsilon
(-p),\quad K(p)=K(-p), \\
& \varepsilon(p)\equiv\int dx\,e^{ipx}\varepsilon(x).
\end{align}
We will now determine the precise form of $K$ by an explicit computation.

\subsubsection{Gravitational part}

\label{3.6.1} In this subsection we will evaluate the gravitational part of
the symplectic form, which is given by the integral of the Crnkovi\'{c}%
-Witten current (\ref{CW}). It should be stressed that splitting the full
symplectic form into the `gravitational part' and `five-form part' has no
physical meaning; in particular, these parts will not be separately
invariant under gauge transformations or changes of $\Sigma$. It is done
here for purely computational purposes.

The background values of the fields characterizing the plane wave metric are
found from (\ref{V}) for $Z=\frac{1}{2}\text{Sign}\,x_{2}$ and are given by (%
$r\equiv\sqrt{x_{2}^{2}+y^{2}}$)%
\begin{equation}
z=\frac{x_{2}}{2r}\,,\quad V_{1}=-\frac{1}{2r},\quad V_{2}=0\,.
\end{equation}
To find the perturbations of these fields, we have to expand (\ref{V}) to
the first order in $\varepsilon(x)$. As we noted above, it is natural to
Fourier-transform w.r.t.\ $x_{1}$. In this basis, the perturbations have the
following form \cite{our}:%
\begin{align}
\delta z(p) & =-y^{2}\frac{1+r|p|}{2r^{3}}e^{-r|p|}\varepsilon (p)\,,  \notag
\\
\delta V_{1}(p) & =-x_{2}\frac{1+r|p|}{2r^{3}}e^{-r|p|}\varepsilon (p)\,,
\notag \\
\delta V_{2}(p) & =\frac{ip}{2r}e^{-r|p|}\varepsilon(p)\,,  \label{FT}
\end{align}
where
\begin{equation*}
\delta z(p)\equiv\delta z(p,x_{2},y)\equiv\int dx_{1}\,e^{ipx_{1}}\delta
z(x_{1},x_{2},y),\quad\text{etc}.
\end{equation*}
To find perturbations of the metric components one should also vary Eqs.\ (%
\ref{hG}) and use the above $\delta z$. We do not present the results of
these elementary computations here.

According to (\ref{integral}), we have to compute the time-component of the
Crnkovi\'{c}-Witten current (\ref{CW}). The computation is straightforward.
The current is a bilinear expression in $\varepsilon(p_{i})$, $i=1,2$. Here
we present the result of integrating this expression in $x_{1}$, which
produces a momentum-conserving $\delta$-function $p_{1}+p_{2}=0$:
\begin{equation}
\int dx_{1}J_{G}^{t}=\int\frac{dp}{2\pi}\left[ 2r^{2}p^{2}+2r|p|-1+\left(
2r^{2}p^{2}+4r|p|+3\right) \frac{x_{2}^{2}}{r^{2}}\right] \frac{ipy^{3}}{%
2r^{3}}e^{-2r|p|}\varepsilon(p)\wedge\varepsilon(-p)\,.
\end{equation}
This expression still has to be integrated in $y>0$ and $x_{2}$. The
dependence on $p$ can be scaled out of the integrand by making the change of
variables $r\rightarrow r/|p|$. The remaining integral is easy to evaluate
in polar coordinates, and the final result is:%
\begin{equation}
\int dx_{1}\,dx_{2}\,dy\,J_{G}^{t}=\int\frac{dp}{2\pi}\frac{i}{p}%
\varepsilon(p)\wedge\varepsilon(-p)\,.  \label{pG}
\end{equation}
This result was also derived in \cite{our} by using the general expressions
for the symplectic current, which we will present in Section \ref{3.7.1}
below.

\subsubsection{Five-form part}

\label{3.6.2} Perturbations of the gauge field potentials can be analogously
found from (\ref{axial}) \cite{our}:
\begin{align}
\delta B_{1}(p) & =-\frac{1}{4}e^{-r|p|}\,\left[ x_{2}|p|+(1+|p|r)\right]
\varepsilon(p)\,,  \notag  \label{1} \\
\delta\tilde{B}_{1}(p) & =-\frac{1}{4}e^{-r|p|}\,\left[ x_{2}|p|-(1+|p|r)%
\right] \varepsilon(p)\,,
\end{align}%
\begin{equation}
\delta B_{2}(p)=(i\,\text{Sign\thinspace}p)\,\delta B_{1}(p)\,,\qquad \delta%
\tilde{B}_{2}(p)=-(i\,\text{Sign\thinspace}p)\,\delta B_{1}(p)\,.  \label{e2}
\end{equation}
As we found in \cite{our}, the bulk term in the gauge field symplectic form
expression (\ref{corrected}) vanishes on these perturbations:%
\begin{equation}
\int d^{2}x\,dy\,J_{F}^{\text{bulk}}=0\qquad\text{(plane wave)\thinspace.}
\label{bulk}
\end{equation}
Thus the only contribution will come from the boundary term in (\ref%
{currents}). The field strengths at $y=0$ are easily found to be%
\begin{align}
\delta F_{12}(p) & =-p^{2}x_{2}e^{-x_{2}|p|}\varepsilon(p)\qquad (x_{2}>0),
\notag \\
\delta\tilde{F}_{12}(p) & =p^{2}x_{2}e^{x_{2}|p|}\varepsilon(p)\qquad
(x_{2}<0).
\end{align}
The field strength variations vanish in the regions complementary to the
given, and thus the corresponding variations of the potentials are pure
gauge, in agreement with our discussion in Section \ref{3.5}. Solving (\ref%
{tosolve}), we find the gauge transformation parameters:%
\begin{align}
\lambda(p) & =-\frac{i}{4p}e^{x_{2}|p|}\varepsilon(p)\,\qquad(x_{2}<0),
\notag \\
\tilde{\lambda}(p) & =\frac{i}{4p}e^{-x_{2}|p|}\varepsilon(p)\,\qquad
(x_{2}>0).
\end{align}
With this information it is possible to evaluate the boundary term in (\ref%
{corrected}):%
\begin{equation}
\int_{y=0}d^{2}x\,J_{F}^{\text{bdry}}\,=8\int_{y=0,x_{2}<0}\,\lambda
\wedge\delta\tilde{F}_{12}-8\int_{y=0,x_{2}>0}\,\tilde{\lambda}\wedge\delta
F_{12}\,=\,\int\frac{dp}{2\pi}\frac{i}{p}\varepsilon(p)\wedge\varepsilon
(-p)\,\,.  \label{pF}
\end{equation}

\subsubsection{Plane wave symplectic form}

\label{3.6.3} Adding the contributions of (\ref{pG}) and (\ref{pF}) in (\ref%
{integral}), we get the symplectic form for perturbations around the plane
wave background:%
\begin{equation}
\omega=\frac{4\pi^{4}}{\kappa_{10}^{2}}\int\frac{dp}{2\pi}\frac{i}{p}%
\,\varepsilon(p)\wedge\varepsilon(-p).
\end{equation}
Since $i/p$ is the Fourier transform of $\frac{1}{2}$Sign$x$, this can be
rewritten in the coordinate space as%
\begin{equation}
\omega=\frac{2\pi^{4}}{\kappa_{10}^{2}}\int\int dx\,d\tilde{x}\,\text{Sign}%
(x-\tilde{x})\,\varepsilon(x)\wedge\varepsilon(\tilde{x}).
\end{equation}
We would now like to compare this answer with the fermion symplectic form $%
\omega_{\text{ferm}}$ given by (\ref{fermform}). These symplectic forms
encode Poisson brackets between different Fourier harmonics of the boundary
curve perturbation $\gamma_{\perp}(s)\equiv\varepsilon(x)$. Strictly
speaking, we must compare commutation relations, which differ from the
Poisson brackets by an extra Planck constant factor, $\hbar_{10}$ on the
supergravity side and $\hbar$ on the fermion side. Since the 10-dimensional
Planck constant satisfies the usual convention $\hbar_{10}=1,$ the precise
relation which should be satisfied is%
\begin{equation}
\omega=\hbar^{-1}\omega_{\text{ferm}}\,.  \label{precise}
\end{equation}
Using the value of $\hbar$ from (\ref{hbar}), we see that this is indeed
true.

\subsection{General droplet case}

\label{3.7} In this section we will compute the symplectic form on the full
moduli space.

\subsubsection{Current expressions}

\label{3.7.1} The first step is to evaluate the symplectic current as a
function of variations of the various fields describing the solutions. The
gravitational part of the symplectic current was evaluated in \cite{our} and
is given by
\begin{align}
J_{G}^{t}=~y^{3}\biggl[ -\frac{1}{4}\delta(V_{i}\partial_{i}h^{-4})\wedge%
\delta(h^{4})+3\delta(V_{i}\partial_{i}G)\wedge\delta G&+2\delta
(h^{-4}V_{j}\,\partial_{\lbrack i}V_{j]})\wedge\delta V_{i}  \notag \\
&-4\delta (\partial_{i}\ln h)\wedge\delta V_{i}\biggr] \,.
\end{align}
The last term can be represented as a total derivative by using $\partial
_{i}V_{i}=0$. In the remaining terms we express $h$ and $G$ via $z$ from (%
\ref{hG}). We also choose to use the linear relations (\ref{linear}) to
eliminate $\partial_{i}z$ and $\partial_{\lbrack i}V_{j]}$ in favor of $%
\partial_{y}V_{i}$ and $\partial_{y}z$. The resulting simplified expression
has the form:
\begin{equation}
J_{G}^{t}=~{\frac{y^{4}\left( \frac{3}{4}+z^{2}\right) }{\left( \frac{1}{4}%
-z^{2}\right) ^{2}}\varepsilon_{ij}\,}\delta(V_{j}\,\partial_{y}V_{i})\wedge%
\delta z-y^{4}\varepsilon_{ij}\,\delta\left( {\frac {V_{i}\,\partial_{y}z}{%
\frac{1}{4}-z^{2}}}\right) \wedge\delta V_{j}-4\partial_{i}\left( \delta\ln
h\wedge\delta V_{i}\right) \,.  \label{JG}
\end{equation}
To compute the five-form symplectic current, it is convenient to rewrite the
gauge potential perturbations (\ref{axial}) in the form%
\begin{align}
\delta B_{i} & =-\frac{1}{4}(a_{i}+b_{i}),\quad\delta\tilde{B}_{i}=-\frac {1%
}{4}(a_{i}-b_{i}),  \notag \\
a_{i} & \equiv\delta\left[ \frac{y^{2}V_{i}}{2\left( {\frac{1}{4}}%
-z^{2}\right) }+U_{i}\right] \,,  \label{conv} \\
b_{i} & \equiv\delta\left[ \frac{y^{2}z\,V_{i}}{{\frac{1}{4}}-z^{2}}\right]
\,.  \notag
\end{align}
The bulk integrand in (\ref{corrected}) then takes the form:%
\begin{align}
J_{F}^{\text{bulk}} &
=\varepsilon_{ij}(a_{i}\wedge\partial_{y}b_{j}-\partial_{y}a_{i}\wedge
b_{j})\,  \notag \\
& =-2\varepsilon_{ij}\,\partial_{y}a_{i}\wedge b_{j}\,+\partial_{y}\left(
\varepsilon_{ij}a_{i}\wedge b_{j}\right) \,  \notag \\
& =-2\varepsilon_{ij}\,\delta\left[ \partial_{y}\frac{y^{2}V_{i}}{2\left( {%
\frac{1}{4}}-z^{2}\right) }-2yV_{i}\right] \wedge\delta\left[ \frac {%
y^{2}z\,V_{j}}{{\frac{1}{4}}-z^{2}}\right] +\partial_{y}\left(
\varepsilon_{ij}a_{i}\wedge b_{j}\right) \,,\,  \label{JF}
\end{align}
where we used (\ref{linear}) to express $\partial_{y}U_{i}$ via $V_{i}$ in
the first term of this expression, so that it became a function of $V_{i}$
and $z$ only.

\subsubsection{Total derivative representation}

\label{3.7.2} To find the symplectic form, we will have to integrate the
bulk currents $J_{G}^{t}$ and $J_{F}^{\text{bulk}}$ found in the previous
section. We see that these currents are given by complicated nonlinear
expressions, so that performing this integration seems to be a daunting
task. Our only hope to do the integral is to look for a total derivative
representation of the total bulk current:%
\begin{equation}
J^{\text{bulk}}\equiv J_{G}^{t}+J_{F}^{\text{bulk}}\overset{?}{=}\partial
_{y}I^{y}+\partial_{i}I^{i}\,.  \label{total}
\end{equation}
If such a representation exists, then the bulk integral will localize to the
(coordinate) boundary and will be computable.

As we will explain now, there is a systematic way to look for
representations of this kind. First of all, notice that the last terms of (%
\ref{JG}) and (\ref{JF}) already have the required total derivative form.
Let $\tilde
{J}^{\text{bulk}}$ denote the sum of the remaining terms. We can
expand $\tilde{J}^{\text{bulk}}$ in the basis of 2-forms built out of $%
\delta z$, $\delta V_{i}$ and their $y$-derivatives:%
\begin{equation}
\tilde{J}^{\text{bulk}}=A\epsilon_{ij}\delta V_{i}\wedge\delta
V_{j}+B\epsilon_{ij}\partial_{y}\delta V_{i}\wedge\delta V_{j}+M_{i}\delta
V_{i}\wedge\partial_{y}\delta z+N_{i}\partial_{y}\delta V_{i}\wedge\delta
z\,+P_{i}\delta V_{i}\wedge\delta z.
\end{equation}
The coefficients in this expansion are easy to find from (\ref{JG}), (\ref%
{JF}), and we won't present them here.

Now we consider the following ansatz:%
\begin{align}
\tilde{J}^{\text{bulk}} & =\partial_{y}\tilde{I}^{y}+\partial_{i}\tilde {I}%
^{i}\,,  \notag \\
\tilde{I}^{y} & =\alpha\,\varepsilon_{ij}\delta V_{i}\wedge\delta
V_{j}+\beta_{i}\,\delta V_{i}\wedge\delta z\,,  \label{ans1} \\
\tilde{I}^{i} & =F\,\delta V_{i}\wedge\delta z\,.  \notag
\end{align}
This ansatz is natural if we assume that $\tilde{I}^{a}$ is given by local
expressions in field variations. Since $\tilde{I}^{a}$ should contain one
derivative less than $\tilde{J}^{\text{bulk}},$ it should be built entirely
from $\delta z$ and $\delta V_{i}$. Comparing the coefficients, we get the
following set of equations:%
\begin{align}
A & =\partial_{y}\alpha  \notag \\
B & =2\alpha-yF  \notag \\
M_{i} & =N_{i}=\beta_{i}  \notag \\
P_{i} & =\partial_{y}\beta_{i}+\partial_{i}F
\end{align}
(to get the second equation, one has to express $\partial_{i}z$ via $%
\partial_{y}V_{i}$ using (\ref{linear})). It turns out that these equations
indeed admit a solution. Namely, from the first three equations one finds:%
\begin{align}
\alpha & =-\frac{y^{4}z\left( \frac{1}{4}+z^{2}\right) }{\left( \frac {1}{4}%
-z^{2}\right) ^{2}}\,,  \notag \\
\beta_{i} & =-{\frac{y^{4}\epsilon_{ij}V_{j}}{\frac{1}{4}-z^{2}}-\frac{%
2y^{4}z^{2}\varepsilon_{ij}V_{j}}{\left( \frac{1}{4}-z^{2}\right) ^{3}}\,,}
\label{sol} \\
F & =\frac{2y^{3}z}{\frac{1}{4}-z^{2}}\,.  \notag
\end{align}
One then can check, using (\ref{linear}), that the remaining fourth equation
is also satisfied, and thus we have constructed a total derivative
representation for $\tilde{J}^{\text{bulk}}$. The total derivative
representation for $J^{\text{bulk}}$ follows when we take the last terms in (%
\ref{JG}) and (\ref{JF}) into account:%
\begin{align}
J^{\text{bulk}} & =\partial_{y}I^{y}+\partial_{i}I^{i}\,,  \notag \\
I^{y} & =\varepsilon_{ij}\delta V_{i}\wedge\delta(\alpha
V_{j})+\varepsilon_{ij}a_{i}\wedge b_{j}\,, \\
I^{i} & =F\,\delta V_{i}\wedge\delta z\,-4\delta\ln h\wedge\delta V_{i}\,.
\notag
\end{align}
Here it was possible to simplify the $I^{y}$ by noticing that the found
coefficients $\alpha$ and $\beta_{i}$ are in fact related by
\begin{equation}
\beta_{i}=\frac{\partial\alpha}{\partial z}\varepsilon_{ij}V_{j}\,.
\end{equation}

We end this subsection with the following remark. One could think that since
the symplectic current is conserved, it can always be represented as a total
derivative, i.e.\ we will always have
\begin{equation}
\ast J=d\Theta\,.
\end{equation}
This is of course true, at least locally. But there is a crucial point,
which makes looking for a representation of this kind rather useless in
general. Namely, while $J$ of course is given by an expression which is
local in the variations of the fields, $\Theta$ in general will be
non-local. For instance, in the chiral boson example considered in Section %
\ref{2.2}, the symplectic current restricted to the chiral boson moduli
space has the form%
\begin{equation}
J^{v}=\delta f^{\prime}(u)\wedge\delta f(u)\,.
\end{equation}
It is obvious that this cannot be represented as a $u$-derivative of an
expression local in $\delta f$.

It is thus clear that something special is going on in our case, since we
did find a total derivative representation local in the variations of the
fields. It would be extremely interesting to provide an explanation for this
phenomenon, and in particular to find out if supersymmetry plays any role in
it.

\subsubsection{Symplectic form}

\label{3.7.3} Using the total derivative representation found in the
previous section, we find that the integral of $J^{\text{bulk}}$ localizes
to the (coordinate) boundary. Since $\delta V_{i}$ decays at $%
x,y\rightarrow\infty$, the only contribution will come from $I^{y}$ and it
is located at $y=0$. Adding the $J_{F}^{\text{bdry}}$ contribution (see (\ref%
{currents})), we get%
\begin{align}
\omega_{0} & \equiv\int
dy\,d^{2}x(J_{G}^{t}+J_{F}^{t})=\int_{y=0}d^{2}x\,(-I^{y}+J_{F}^{\text{bdry}%
})  \notag \\
& =\int_{y=0}\,\left[ -\,\varepsilon_{ij}\delta V_{i}\wedge\delta(\alpha
V_{j})\,-\varepsilon_{ij}a_{i}\wedge b_{j}\,+8\,(\lambda\wedge\delta\tilde {F%
}_{12}-\tilde{\lambda}\wedge\delta F_{12}\,)\right] \,\,.  \label{o0}
\end{align}
We now proceed with further simplifications, beginning with the last term.
We have%
\begin{align}
\int_{y=0}\,\,(\lambda\wedge\delta\tilde{F}_{12}&-\tilde{\lambda}%
\wedge\delta F_{12}\,) =\int_{\mathcal{D}}\,\lambda\,\wedge\varepsilon^{ij}%
\,\partial_{i}\delta\tilde{B}_{j}-\int_{\mathcal{D}^{c}}\,\tilde{\lambda }%
\,\wedge\varepsilon^{ij}\,\partial_{i}\delta B_{j}  \notag \\
& =-\int_{\mathcal{D}}\,\,\varepsilon^{ij}\,\partial_{i}\lambda\wedge \delta%
\tilde{B}_{j}+\int_{\mathcal{D}^{c}}\,\varepsilon^{ij}\,\partial _{i}\tilde{%
\lambda}\,\wedge\delta B_{j}+\int_{\partial\mathcal{D}}(\lambda\wedge\delta%
\tilde{B}^{-}+\tilde{\lambda}\wedge\delta B^{+})  \notag \\
& =\int_{y=0}(2z)\,\,\varepsilon^{ij}\delta B_{i}\wedge\delta\tilde{B}%
_{j}+\int_{\partial\mathcal{D}}(\lambda\wedge\delta\tilde{B}^{-}+\tilde{%
\lambda}\wedge\delta B^{+})
\end{align}
Here in the first line we used the fact, following from the general
discussion in Section \ref{3.5}, that $\delta F_{12}(\delta\tilde{F}_{12})$
will vanish in $\mathcal{D}(\mathcal{D}^{c})$. In the second line we used
Stokes' theorem, and in the third line we used (\ref{tosolve}) and also
united the integrals over $\mathcal{D}$ and $\mathcal{D}^{c}$ by inserting
the $2z$ factor to account for the sign difference. The superscript $+(-)$
on $\delta\tilde{B}$ and $\delta B$ indicates that the limiting boundary
value on $\partial \mathcal{D}$ should be taken from outside (inside) $%
\mathcal{D}$.

To simplify further, we will need the following expressions, which are easy
to compute from (\ref{conv}):%
\begin{equation}
\varepsilon_{ij}a_{i}\wedge b_{j}=\varepsilon_{ij}\delta V_{i}\wedge \delta
\left[ \frac{y^{4}z\,V_{j}}{2\left( \frac{1}{4}-z^{2}\right) ^{2}}\right]
+\varepsilon_{ij}\delta U_{i}\wedge\delta\left[ \frac{y^{2}z\,V_{j}}{\frac{1%
}{4}-z^{2}}\right] \,,  \label{exp1}
\end{equation}%
\begin{align}
16z\,\varepsilon^{ij}\delta B_{i}\wedge\delta\tilde{B}_{j}=\varepsilon
_{ij}\delta V_{i}\wedge\delta\left[ \frac{y^{4}V_{j}}{\frac{1}{4}-z^{2}}%
\right] & +z\,\varepsilon_{ij}\delta U_{i}\wedge\delta\left[ \frac {%
y^{2}V_{j}}{\frac{1}{4}-z^{2}}\right]  \notag \\
& \qquad-\frac{y^{4}\varepsilon_{ij}V_{j}}{\frac{1}{4}-z^{2}}\,\delta
V_{i}\wedge\,\delta z+z\varepsilon_{ij}\delta U_{i}\wedge\delta U_{j}\,.
\label{exp2}
\end{align}
Plugging these into (\ref{o0}), we get the following simplified expression
for $\omega_{0}$:%
\begin{equation}
\omega_{0}=-\int_{y=0}\frac{y^{2}\varepsilon_{ij}\,V_{j}}{\frac{1}{4}-z^{2}}%
\left( \delta U_{i}+y^{2}\delta V_{i}\right) \wedge\delta z+\int
_{y=0}z\varepsilon_{ij}\delta U_{i}\wedge\delta U_{j}+8\int_{\partial
\mathcal{D}}(\lambda\wedge\delta\tilde{B}^{-}+\tilde{\lambda}\wedge\delta
B^{+}).  \label{sim1}
\end{equation}
What happened here is that the first terms of (\ref{exp1}), (\ref{exp2})
have completely cancelled against $I^{y}$. The second terms of (\ref{exp1}),
(\ref{exp2}) also partially cancelled between themselves.

The first term in (\ref{sim1}) still looks rather complicated, however it
turns out that it vanishes. Roughly, this happens because $\delta z$ for $%
y\rightarrow0$ tends to a $\delta$-function supported on $\partial \mathcal{D%
}.$ At the same time the expression multiplying $\delta z$ tends to a
function which vanishes on $\partial\mathcal{D},$ and thus the integral of
the product tends to zero. A careful demonstration of this fact is given in
Appendix \ref{A.1}.

Now that we are left with just two terms, let us take a close look at the
gauge parameters $\lambda$, $\tilde{\lambda}$. Recall that they have to be
determined on $\mathcal{D}(\mathcal{D}^{c})$ by solving (\ref{tosolve}). It
is easy to see that the first terms of the axial gauge expressions (\ref%
{axial}) vanish in the corresponding domains, and thus we have:%
\begin{align}
\delta B_{i} & =-\frac{1}{4}\delta U_{i}\text{\quad on }\mathcal{D}\,,
\notag \\
\delta\tilde{B}_{i} & =-\frac{1}{4}\delta U_{i}\text{\quad on }\mathcal{D}%
^{c}\,.  \label{vars1}
\end{align}
Thus (\ref{tosolve}) takes the form
\begin{align}
-\frac{1}{4}\delta U_{i} & =\partial_{i}\lambda\quad\text{on }\mathcal{D\,},
\notag \\
-\frac{1}{4}\delta U_{i} & =\partial_{i}\tilde{\lambda}\quad\text{on }%
\mathcal{D}^{c}\,.  \label{tosolve1}
\end{align}
As we explain in Appendix \ref{A.2}, $\delta U|_{y=0}$ is closed and,
moreover, has zero integral on all topologically nontrivial cycles in $%
\mathcal{D}$ and $\mathcal{D}^{c}$. Thus equations (\ref{tosolve1}) always
have a solution.

Using (\ref{tosolve1}), the second term in (\ref{sim1}) can be transformed
into an integral over $\partial\mathcal{D}$:%
\begin{align}
\int_{y=0}z\varepsilon_{ij}\delta U_{i}\wedge\delta U_{j} & =\frac{1}{2}%
\int_{\mathcal{D}^{c}}\varepsilon_{ij}\delta U_{i}\wedge\delta U_{j}-\frac {1%
}{2}\int_{\mathcal{D}}\varepsilon_{ij}\delta U_{i}\wedge\delta U_{j}  \notag
\\
& =-2\int_{\mathcal{D}^{c}}\varepsilon_{ij}\partial_{i}\tilde{\lambda}%
\wedge\delta U_{j}+2\int_{\mathcal{D}}\varepsilon_{ij}\partial_{i}\lambda%
\wedge\delta U_{j}  \notag \\
& =2\int_{\partial\mathcal{D}}(\tilde{\lambda}\wedge\delta
U^{+}+\lambda\wedge\delta U^{-})\,\,.  \label{term2}
\end{align}
Thus we are reduced to the following representation:
\begin{equation}
\omega_{0}=\int_{\partial\mathcal{D}}\lambda\wedge(2\delta U^{-}+8\delta
\tilde{B}^{-})+\tilde{\lambda}\wedge(2\delta U^{+}+8\delta B^{+})  \label{bb}
\end{equation}
As we explain in Appendix \ref{A.3}, $\delta B$ and $\delta\tilde{B}$ are
continuous across $\partial\mathcal{D}$. Using this information and (\ref%
{vars1}), we have%
\begin{align}
\delta B^{+} & =\delta B^{-}=-\frac{1}{4}\delta U^{-},\qquad  \notag \\
\delta\tilde{B}^{-} & =\delta B^{+}=-\frac{1}{4}\delta U^{+}\,.  \label{b0}
\end{align}
On the other hand, the tangential component $\delta U_{\Vert}$ has a jump
discontinuity on $\partial\mathcal{D},$ the opposite-side limits being given
by (see Appendix \ref{A.3})%
\begin{equation}
\delta U_{\Vert}^{\pm}=\mp\delta\gamma_{\perp}\,.  \label{b1}
\end{equation}
In view of (\ref{tosolve1}), we can express $\lambda,\tilde{\lambda }%
|_{\partial\mathcal{D}}$ via $\delta U_{\Vert}|_{\partial\mathcal{D}}$ as
follows\footnote{%
The choice of the lower limit of these integrals is unimportant and does not
affect the value of the symplectic form below.}%
\begin{align}
\lambda(s) & =-\frac{1}{4}\int_{0}^{s}d\tilde{s}\,\delta U_{\Vert}^{-}(%
\tilde{s})=-\frac{1}{4}\int_{0}^{s}d\tilde{s}\,\delta\gamma_{\perp }(\tilde{s%
})\,,  \notag \\
\tilde{\lambda}(s) & =-\frac{1}{4}\int_{0}^{s}d\tilde{s}\,\delta U_{\Vert
}^{+}(\tilde{s})=\frac{1}{4}\int_{0}^{s}d\tilde{s}\,\delta\gamma_{\perp }(%
\tilde{s})\,.  \label{b2}
\end{align}

Eqs. (\ref{b0}), (\ref{b1}), (\ref{b2}) contain all information necessary to
evaluate (\ref{bb}) in terms of $\delta\gamma_{\perp}.$ We obtain:%
\begin{align}
\omega_{0} & =2\oint ds\,\delta\gamma_{\perp}(s)\wedge\left( \int_{0}^{s}d%
\tilde{s}\,\delta\gamma_{\perp}(\tilde{s})\right) =2\oint\oint ds\,d\tilde{s}%
\,\theta(s-\tilde{s})\,\delta\gamma_{\perp}(s)\wedge\delta \gamma_{\perp}(%
\tilde{s})  \notag \\
& =\oint\oint ds\,d\tilde{s}\,\text{Sign}(s-\tilde{s})\,\delta\gamma_{\perp
}(s)\wedge\delta\gamma_{\perp}(\tilde{s})\,,  \label{full}
\end{align}
where we took advantage of the constraint (\ref{constr}), which allows us to
add a constant to the kernel without changing the symplectic form.

We see that $\omega_{0}$ has exactly the same functional dependence as the
fermion symplectic form $\omega_{\text{ferm}}$ from Section \ref{3.3}. The
supergravity symplectic form $\omega$ differs from $\omega_{0}$ by a
constant factor (see (\ref{integral}), (\ref{o0})). As discussed in Section %
\ref{3.6.3}, the relative normalization of $\omega$ and $\omega_{\text{ferm }%
}$should also match. It is easy to see that the precise matching condition
as expressed by Eq.\ (\ref{precise}) is indeed satisfied, just like in the
plane wave case.

To close the discussion, we would like to note that Eq.\ (\ref{bb}) is
completely general and in particular applies when the droplet boundary has
several disconnected components. In this case we can still use Eq.\ (\ref{b0}%
)-(\ref{b2}) on each boundary component separately, and reproduce the
general fermionic symplectic form (\ref{genform}).

\section{Conclusions and future directions}

\label{concl}

In this paper we have laid out a general procedure for quantizing moduli
spaces of solutions of supergravity, on which all fluxes are kept fixed.
This procedure, the 'on-shell quantization', entails writing down the CWZ
symplectic current for the SUGRA action, evaluating it on the moduli space
and integrating it over a hypersurface to obtain the symplectic form on the
moduli space. The Poisson brackets which are obtained from this symplectic
form are then promoted to commutators, which are used to build the whole
Hilbert space.

We have stressed that in order to obtain a non-degenerate symplectic form,
it is important that the solutions be dynamical (i.e.\ not static). We also
demonstrated that the on-shell quantization is really equivalent to the more
conventional method of expanding the action to second order around a
solution, but has the advantage that it automatically gives the truncation
of all fluctuation modes into the moduli space of interest. We discussed the
corrections to this quantization, appearing upon introducing the full set of
possible fluctuations including those transverse to the moduli space, and
explained that although in general energy levels could shift, the spectrum
may be protected against such corrections in presence of supersymmetry.

As an application of the developed procedure, we analyzed the
family of ``Bubbling AdS" SUGRA solutions found by LLM \cite{LLM}.
We noted that one must work in a gauge choice where the variations
of the metric and RR gauge fields are regular, and wrote down the
expressions for the symplectic current taking this into account.
Remarkably, it turned out the symplectic current around \emph{any
droplet or collection of droplets} can be written as a total
derivative of a local expression. This enabled us to integrate it
and find the symplectic form in full generality, Eq. (\ref{full}).
Comparing this to what we expect from the fermion picture, we find
\emph{exact agreement}.

In other words, we have proved that the Hilbert space around any point in
the moduli space of LLM solutions is the same as that of free fermions in a
harmonic oscillator, both quantized around the same droplet configuration.
We stress again that we only used the SUGRA action and solutions, and in no
way used AdS/CFT or string theory to obtain this result. Thus one can view
this as a derivation of AdS/CFT in the $N\rightarrow\infty$ limit, in this
restricted subsector of 1/2 BPS states\footnote{%
We should stress that we make no claim about the matching of correlation
functions with intermediate states which are not in this sector.}.

In this sense, our on-shell quantization is a powerful tool, especially in
cases where a dual description of a SUGRA system is not known. It some
situations it can be used to derive such a dual picture, where the
microscopic degrees of freedom are manifest. It is also very useful as it
enables one to count the dimensions of certain subsectors of the Hilbert
space --- certain ensembles of microscopic states with common macroscopic
characteristics --- and thus deduce their entropy. This is of course a very
interesting question, especially when one deals with black hole geometries,
such as the 2-charge D1-D5 and 3-charge D1-D5-P black holes that we have
discussed in section \ref{2.1}. In future work, we intend to apply this
method to quantize the 2-charge D1-D5 system \cite{D1D5}. It would be nice
to explore other systems and setups where the on-shell quantization method
could be applied. In particular it would be interesting to understand the
subtleties in applying our method to spacetimes with horizons and to
spacetimes which include both electric and magnetic sources\footnote{%
Recent work \cite{deWit} may be useful in this last respect.}.

Another very interesting question to explore is the $1/N$ corrections to our
results. As we mentioned in Sections \ref{corr} and \ref{3.3}, such
corrections should effectively reduce the dimension of the Hilbert space by
decreasing degeneracies of the energy levels above $N$. It would be
extremely interesting to understand how such a reduction may occur in a
perturbative fashion\footnote{%
An analysis of the fermion system for finite $N$ can be found in \cite{Dhar}%
, where the noncommutative nature of the fermion Wigner density appears.}.

\acknowledgments

We would like to thank D.~Belov, I.~Bena, J.~de Boer, A.~Dhar, S.~Hartnoll,
F.~Larsen, J.~Maldacena, G.~Mandal, D.~Martelli, A.~Naqvi, V.~Schomerus,
K.~Skenderis, N.~Suryanarayana, M.~Taylor, T.~Wiseman and B.~Zwiebach for
useful comments and discussions. VR would like to thank the organizers of
the String Theory 2005 workshop at the Benasque Center for Sciences for
creating a pleasant environment while this work was being completed. This
research was supported by Stichting FOM.

\appendix

\section{Appendix}

\subsection{Fermion symplectic form for multiple-valued $r(\protect\phi)$}

\label{A.0}

Let us show that the Poisson bracket (\ref{bg1}) generates the correct
equations of motion of the droplet boundary for multiple-valued $r(\phi).$
The energy of such droplets is still given by Eq.\ (\ref{ham}), which should
be understood as a line integral along the boundary, so that $d\phi
=(d\phi/ds)ds,$ and $d\phi/ds$ changes sign at points separating different
branches of $r(\phi)$. Taking this change of sign into account, the energy
can be rewritten in terms of separate branches $r=r_{i}(\phi)$ as%
\begin{equation}
H_{\text{tot}}=\,\frac{1}{16\pi\hbar^{2}}\sum_{i}\varepsilon_{i}%
\int_{I_{i}}d\phi\,r_{i}^{4}(\phi),\quad\varepsilon_{i}=\text{Sign}\frac{%
d\phi}{ds}\,,  \label{ham1}
\end{equation}
where $I_{i}$ are the intervals on which the corresponding branches are
defined. Furthermore, each branch satisfies the same equation of motion (\ref%
{chiral}) as before:%
\begin{equation}
\dot{r}_{i}=\hbar^{-1}r_{i}^{\prime}.  \label{chiral1}
\end{equation}

To rewrite (\ref{bg1}) in terms of $r_{i}$, note that the relation (\ref%
{elem}) remains true for each branch (when $s$ crosses from one branch to
the next, both $d\phi /ds$ and $\delta r/\delta \gamma _{\perp }$ in (\ref%
{elem}) change sign). The $\delta ^{\prime }$ transforms as
\begin{equation}
\delta ^{\prime }(s(\phi )-s(\tilde{\phi}))=\frac{\varepsilon _{i}}{%
s^{\prime }(\phi )s^{\prime }(\tilde{\phi})}\delta ^{\prime }(\phi -\tilde{%
\phi})\,,
\end{equation}%
and we get:%
\begin{equation}
\{\delta \,r_{i}^{2}(\phi ),\delta \,r_{j}^{2}(\tilde{\phi})\}=8\pi \hbar
\,\delta _{ij}\varepsilon _{i}\,\delta ^{\prime }(\phi -\tilde{\phi})\,.
\label{ourbr}
\end{equation}%
The Hamilton equations corresponding to this bracket and (\ref{ham1}) indeed
coincide with (\ref{chiral1}).

\subsection{Regular gauge field perturbations}

\label{regul}

In this section we demonstrate the existence and properties of a regular
gauge potential perturbation $\delta A_{4}^{\text{reg}}$ used in Section \ref%
{3.5}. First, we note that the closed 5-form $\delta F$ is regular, since it
is a difference between two regular 5-forms corresponding to neighboring
LLM geometries. Moreover, since we assume that all the fluxes are kept
fixed, $\delta F$ has zero integral on all closed cycles. Thus by de Rahm's
theorem, it is in the trivial cohomology class: there exists a regular
4-form $\delta A_{4}^{\text{reg}}$ so that
\begin{equation}
\delta F=d(\delta A_{4}^{\text{reg}})\,.  \label{fa}
\end{equation}%
Furthermore, this 4-form may be assumed to be $SO(4)\times SO(4)$ invariant,
and thus of the form (\ref{dAreg}). This can be achieved by averaging over
this group. More precisely, every group element $g\in SO(4)\times SO(4)$
acts in an obvious fashion on the LLM manifold, rotating the spheres $S^{3}$
and $\tilde{S}^{3}$. This action leaves $\delta F$ invariant: $g^{\ast
}(\delta F)=\delta F$. Since the pull-back action commutes with exterior
differentiation, we get%
\begin{equation}
\delta F=d\left( g^{\ast }(\delta A_{4}^{\text{reg}})\right) \,.
\end{equation}%
Averaging this last equation over $g$, we get the desired $SO(4)\times SO(4)$%
-invariant representation.

Finally, let us discuss the requirements imposed on the components of $%
\delta B^{\text{reg}},\delta \tilde{B}^{\text{reg}}$ by the regularity of (%
\ref{dAreg}). Consider the case of $\delta B^{\text{reg}}$ first.
Although the sphere $S^{3}$ shrinks to zero size on $\mathcal{D}$,
it does so in a regular fashion, uniting with the $y$ coordinate
to form a metric which
looks locally like the origin of $\mathbb{R}^{4}$\cite{LLM}:%
\begin{equation}
c(x)(dy^{2}+y^{2}d\Omega _{3}^{2})\,.
\end{equation}%
The volume form of this $\mathbb{R}^{4}$, which is of course a
regular 4-form, looks in these coordinates like $y^{3}dy\wedge
d\Omega _{3}$.
Comparing this with (\ref{dAreg}), we see that in order for $\delta A_{4}^{%
\text{reg}}$ to be regular as $y\rightarrow 0$, $\delta B_{y}^{\text{reg}}$
should decay in this limit at least just as fast:%
\begin{equation}
\delta B_{y}^{\text{reg}}=\mathcal{O}(y^{3})\quad \text{on }\mathcal{D}.
\end{equation}%
To get a condition for the decay of $\delta B_{i}^{\text{reg}},\delta B_{t}^{%
\text{reg}}$, we should find a minimal power of $y$ which, multiplying $%
d\Omega _{3}$, turns it into a regular 3-form (notice that the $h$
stays finite as $y\rightarrow 0$, and the metric in the $t$ and
$x^{i}$ directions is
completely regular). In terms of the Cartesian $\mathbb{R}^{4}$ coordinates $%
X^{\mu }$, $\mu =1,\ldots ,4$, we have
\begin{equation}
d\Omega _{3}\sim (X^{2})^{-2}\varepsilon ^{\mu \nu \lambda \sigma }X_{\mu
}\,dX_{\nu }\wedge dX_{\lambda }\wedge dX_{\sigma }\,.
\end{equation}%
Thus it is clear that such a minimal regular multiple is
$y^{4}d\Omega _{3}$, and the remaining regularity condition is:
\begin{equation*}
\delta B_{i}^{\text{reg}},\delta B_{t}^{\text{reg}}=\mathcal{O}(y^{4})\quad
\text{on }\mathcal{D}.
\end{equation*}%
The case of $\delta \tilde{B}^{\text{reg}}$ is completely analogous, and the
regularity conditions are
\begin{equation}
\delta \tilde{B}_{i}^{\text{reg}},\delta \tilde{B}_{t}^{\text{reg}}=\mathcal{%
O}(y^{4}),\quad \delta \tilde{B}_{y}^{\text{reg}}=\mathcal{O}(y^{3})\quad
\text{on }\mathcal{D}^{c}\,.
\end{equation}%
In fact, only a weak form (\ref{reg}) of the above conditions were used in
Section \ref{3.5}.

\subsection{A vanishing integral}

\label{A.1} 
\begin{figure}[tb]
\begin{center}
\epsfig{figure=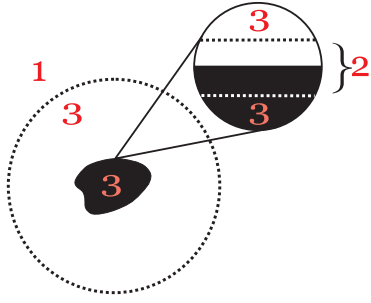,width=5cm}
\end{center}
\caption{For (a collection of) droplets of characteristic size $\ell$, we
split the plane into three regions: the near-infinity region $R_{1}=\{x:%
\text{dist}(x,\mathcal{D})\gg\ell\},$ the
near-$\partial\mathcal{D}$ region
$R_{2}=\{x:\text{dist}(x,\partial\mathcal{D})\ll\ell\},$ and the
intermediate region $R_{3}$ covering the rest of the plane. To
show that
integral (\protect\ref{toshow}) vanishes, we use large-$x$ asymptotics in $%
R_{1}$, wavy line approximation in $R_{2}$, and uniform convergence in $%
R_{3} $.}
\label{vanishpic}
\end{figure}

In this section we show that the first term in (\ref{sim1})
vanishes:
\begin{equation}
\int_{y=0}d^{2}x\,\frac{y^{2}\varepsilon_{ij}\,V_{j}}{\frac{1}{4}-z^{2}}%
\left( \delta U_{i}+y^{2}\delta V_{i}\right) \wedge\delta z=0  \label{toshow}
\end{equation}
The integral here should be understood as the $\bar{y}\rightarrow0$ limit of
an integral over the plane at $y=\bar{y}$. We will assume that $\mathcal{D}$
is a collection of finite-size droplets, and that the boundary $\partial
\mathcal{D}$ consists of one or more smooth closed curves. We will carry out
the analysis separately for the following three regions of the $x_{1},x_{2}$
plane, depending on the distance from $\partial\mathcal{D}$: near-infinity
region $R_{1}$, near-$\partial\mathcal{D}$ region $R_{2}$, and an
intermediate region $R_{3}$ (see Fig.\ \ref{vanishpic}).

The region $R_{3}$ is the simplest, since here $\delta
z=\mathcal{O}(y^{2})$ tends uniformly to zero, while the remaining
factors tend uniformly to a
finite limit. Thus the integral over this bounded region vanishes as $%
\mathcal{O}(\bar{y}^{2})$.

The region $R_{1}$ consists of points $x$ with $|x|$ much larger than the
characteristic size of $\mathcal{D}$. In this case a more careful analysis
is needed, since we must check that the integral converges near infinity.
First we estimate $z$ by
\begin{equation}
z=\frac{1}{2}-\frac{1}{\pi}\frac{y^{2}}{(x^{2}+y^{2})^{2}}\ast\chi _{%
\mathcal{D}}<\frac{1}{2}-\frac{1}{4\pi}\frac{y^{2}}{x^{4}}\ast\chi _{\mathcal{%
D}}\,, \label{Az}
\end{equation}
where $\chi_{\mathcal{D}}$ is the characteristic function, equal $1$ on $%
\mathcal{D}$ and $0$ outside. This estimate is true as long as $y$
is smaller than $\text{dist}(x,\mathcal{D})$, and thus applies in
$R_1$ for $y\lesssim \ell$. From this estimate we get
\begin{equation}
\frac{y^{2}}{\frac{1}{4}-z^{2}}\lesssim\left( \frac{1}{x^{4}}\ast \chi_{%
\mathcal{D}}\right) ^{-1}=\mathcal{O}(x^{4})\,.
\end{equation}
Further, we have%
\begin{align}
|V_{i}| & \lesssim\frac{|x|}{(x^{2}+y^{2})^{2}}\ast\chi_{\mathcal{D}}\,=%
\mathcal{O}(|x|^{-3})\,,  \notag \\
|U_{i}| & \lesssim\frac{|x|}{x^{2}+y^{2}}\ast\chi_{\mathcal{D}}\,=\mathcal{O}%
(|x|^{-1})\,.  \label{AVU}
\end{align}
Finally, from (\ref{AVU}) and (\ref{Az}) we can also conclude that
\begin{equation}
\delta V_{i}=\mathcal{O}(|x|^{-3})\,\,,\quad\delta U_{i}=\mathcal{O}%
(|x|^{-1})\,,\quad\delta z=\mathcal{O}(y^{2}|x|^{-4})\,.
\end{equation}
Combining all these estimates, we see that the integrand in (\ref{toshow})
is $\mathcal{O}(y^{2}|x|^{-4})$. Thus the integral over $R_{1}$ is
convergent and tends to zero as $\mathcal{O}(\bar{y}^{2}).$

Finally consider the region $R_{2}$, consisting of points $x$ which are much
closer to $\partial\mathcal{D}$ than the droplet size. In this region we can
neglect the curvature of the boundary. Without loss of generality, we will
orient the axes so that the droplet covers the lower half-plane (up to
higher-order terms). The situation now becomes identical to the wavy line
approximation used in Section \ref{3.6}. Borrowing results from that
section, we obtain the following estimates for various factors in (\ref%
{toshow}):%
\begin{align}
\frac{y^{2}}{\frac{1}{4}-z^{2}} & =\mathcal{O}(r^{2})\,,\qquad V_{i}=%
\mathcal{O}(r^{-1})\,,  \notag \\
\delta V_{i} & =\mathcal{O}(r^{-2})\,,\qquad\delta U_{i}=\mathcal{O}(1)\,,
\label{various} \\
\delta z & =\mathcal{O}(y^{2}r^{-3})\,.  \notag
\end{align}
For example, the last of these estimates is obtained from (\ref{FT}) as
follows:%
\begin{align}
|\delta z(x_{1},x_{2},y)| & =\frac{y^{2}}{2r^{3}}\left\vert \int\frac {dp}{%
2\pi}e^{-ipx_{1}}(1+r|p|)e^{-r|p|}\varepsilon(p)\right\vert \,  \notag \\
& \leq\frac{y^{2}}{2r^{3}}\max_{p}\left[ \left( 1+r|p|\right) e^{-r|p|}%
\right] \int\frac{dp}{2\pi}|\varepsilon(p)|\,,
\end{align}
which gives us the required estimate under the natural assumption that the
perturbation $\varepsilon(x)$ is smooth, so that the Fourier tranform $%
\varepsilon(p)$ is integrable. The estimates for $\delta V_{i}$ and $\delta
U_{i}$ can be proved in exactly the same way. In the case of $\delta U_{i}$ one
should use the following Fourier transform expressions (see \cite{our},
Appendix B)%
\begin{align}
\delta U_{1}(p) & =\frac{x_{2}}{r}e^{-r|p|}\varepsilon(p)\,,  \notag \\
\delta U_{2}(p) & =i(\text{Sign}p)e^{-r|p|}\varepsilon(p)\,.  \label{UFT}
\end{align}

Using (\ref{various}), we see that the integrand of (\ref{toshow}) is $%
\mathcal{O}(y^{2}r^{-2}).$ This gives the following upper bound for the
integral over $R_{2}$:%
\begin{equation}
\oint_{\partial\mathcal{D}}ds\int_{-\infty}^{\infty}dx_{\perp}\frac{y^{2}}{%
x_{\perp}^{2}+y^{2}}=\mathcal{O}(y),
\end{equation}
and we conclude that this part of (\ref{toshow}) also vanishes as $\bar
{y}%
\rightarrow0.$ The proof of (\ref{toshow}) is now complete.

\subsection{Existence of solutions for $\protect\lambda$ and $\tilde{\protect%
\lambda}$}

\label{A.2} In this section we prove that (\ref{tosolve1}) always admits a
solution. To show this, we use the following explicit form of $\delta
U|_{y=0}$, which follows directly from (\ref{U}):%
\begin{align}
\delta U_{i}(x) & =\frac{1}{\pi}\sum_{b=1}^{B}\oint_{\gamma^{(b)}}K_{i}%
\left( x-\gamma^{(b)}(s)\right) \delta\gamma_{\perp}^{(b)}(s)\,ds\,,  \notag
\\
K_{i}(x) & =\frac{\varepsilon_{ij}x_{j}}{x^{2}}\,.  \label{Uexpl}
\end{align}
Here we use the same notation as in (\ref{genform}) to parametrize $\partial%
\mathcal{D}$ and its pertubations, which are as usual assumed to satisfy the
constraints (\ref{own}).

Now we observe that the convolution kernel in (\ref{Uexpl}) is closed away
from the origin:%
\begin{equation}
\varepsilon_{ij}\partial_{i}K_{j}=\partial_{i}\frac{x_{i}}{x^{2}}%
=2\pi\,\delta^{(2)}(x)\,.  \label{df}
\end{equation}
From this it follows that the integral of $K$ over a closed contour $\Gamma$
depends only on its winding number $n$ around the origin:%
\begin{equation}
\oint_{\Gamma}K=2\pi n\,.\,
\end{equation}
Using this in (\ref{Uexpl}), we get%
\begin{equation}
\oint_{\Gamma}\delta U=0  \label{Ucl}
\end{equation}
for any closed contour lying entirely in $\mathcal{D}$ or $\mathcal{D}^{c}.$
Indeed, such a contour will have definite winding numbers $n^{(b)}$ around
every component of $\partial\mathcal{D}$. The integral in (\ref{Ucl}) will
be given by a linear combination of constraints (\ref{own}) with these
winding numbers as coefficients, and thus vanishes.

\begin{figure}[tb]
\centering {\ \epsfig{figure=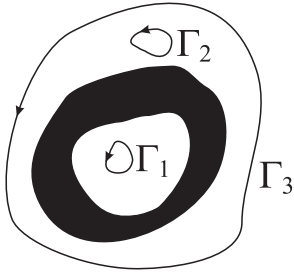,width=4cm} }
\caption{The vector field $\protect\delta U$ has vanishing integrals over
topologically trivial ($\Gamma_{1,2}$) as well as nontrivial ($\Gamma_{3}$)
contours in the complement of $\partial\mathcal{D}$. This is enough to show
that (\protect\ref{tosolve1}) has a global solution.}
\label{contours}
\end{figure}

Eq.\ (\ref{Ucl}) applied to arbitrary topologically trivial contours $\Gamma$
implies that $\delta U$ is closed, which guarantees that (\ref{tosolve1})
has a solution locally. Applied to topologically nontrivial contours, (\ref%
{Ucl}) shows that this local solution has trivial monodromy and is thus
global, which concludes the argument (see Fig.~\ref{contours}).

\subsection{Limits on $\partial\mathcal{D}$}

\label{A.3}

It is obvious from (\ref{axial}), (\ref{U}) that $B,\tilde{B},U$ stay finite
when $y\rightarrow0$, at least when $x\notin\partial D$. Consequently, their
perturbations $\delta B,\delta\tilde{B},\delta U$ will also have finite $y=0$
limits. In this section we study how these latter limiting values behave on $%
\partial\mathcal{D}$. Similarly to how we analyzed region $R_{3}$ in
Appendix \ref{A.1}, we can neglect the curvature of the boundary in the
limit $x\rightarrow\partial\mathcal{D}$, and use the wavy line approximation
of Section \ref{3.6}.

The Fourier transforms of $\delta B_{i}$ and $\delta\tilde{B}_{i}$ with
respect to the longitudinal variable $x_{1}$ are given by (\ref{e2}), where
we have to put $y=0.$ These Fourier transforms are manifestly continuous at $%
x_{2}=0.$ Thus we conclude that $\delta B_{i}$ and $\delta\tilde{B}_{i}$ are
continuous across $\partial\mathcal{D}.$

The Fourier transforms of $\delta U_{i}$ are given by (\ref{UFT}). Setting $%
y=0,$ we see that $\delta U_{2}$ is continuous at $x_{2}=0$, while $\delta
U_{1}$ has a jump discontinuity:%
\begin{equation}
\delta U_{1}^{\pm}=\pm\varepsilon\,.
\end{equation}
This is in agreement with (\ref{b2}), since $\delta U_{\Vert}=-\delta U_{1}$
corresponds to the positive orientation of the droplet boundary.

\end{document}